\documentclass[final]{elsarticle}

\usepackage{etoolbox}
\makeatletter
\patchcmd{\ps@pprintTitle}
{Preprint submitted to}
{Accepted version in}
{}{}
\makeatother

\usepackage{graphicx}
\usepackage{amsmath}
\usepackage{amssymb}
\usepackage{lineno}

\usepackage{epstopdf}
\usepackage[linesnumbered, ruled]{algorithm2e}
\usepackage{nomencl}
\makenomenclature

\makeindex

\journal{Acta Astronautica}

\begin{document}
	
\begin{frontmatter}
\title{A Flatness-Based Predictive Controller for Six-Degrees of Freedom Spacecraft Rendezvous}
\author[1]{Julio C. Sanchez} 
\author[1]{Francisco Gavilan}
\author[1]{Rafael Vazquez\corref{cor1}}
\ead{rvazquez1@us.es}
\author[2]{Christophe Louembet}
\address[1]{Departamento de Ingeniería Aeroespacial, Escuela Técnica Superior de Ingeniería, Universidad de Sevilla, 41092, Sevilla, Spain}
\address[2]{LAAS-CNRS, Université de Toulouse, CNRS, 31400, Toulouse, France}

\cortext[cor1]{Corresponding author}


\begin{abstract}
This work presents a closed-loop guidance algorithm for six-degrees of freedom spacecraft rendezvous with a passive target flying in an eccentric orbit. The main assumption is that the chaser vehicle has an attitude control system, based on reaction wheels, providing the necessary torque to change its orientation whereas the number of thrusters is arbitrary. The goal is to design fuel optimal manoeuvres while satisfying operational constraints and rejecting disturbances. The proposed method is as follows; first, the coupled translational and angular dynamics are transformed to equivalent algebraic relations using the relative translational states transition matrix and the attitude flatness property. Then, a direct transcription method, based on B-splines parameterization and discretization of time continuous constraints, is developed to obtain a tractable static program. Finally, a Model Predictive Controller, based on linearization around the previously computed solution, is considered to handle disturbances. Numerical results are shown and discussed.
\end{abstract}
\end{frontmatter}

\makenomenclature
\nomenclature[1]{$\mathbf{B}$}{Input matrix}
\nomenclature[2]{$B$}{Chaser body frame}
\nomenclature[3]{$c_y$, $c_z$}{LOS region parameters}
\nomenclature[4]{\text{E}}{Mathematical expectation}
\nomenclature[5]{$\mathbf{e}$}{Rotation axis}
\nomenclature[6]{$e$}{Target eccentricity}
\nomenclature[7]{$\mathbf{H}$}{Angular momentum}
\nomenclature[8]{$\mathbf{I}$}{Chaser inertia matrix}
\nomenclature[9]{$I$}{Inertial geocentric frame}
\nomenclature[a1]{$\mathbf{Id}$}{Identity matrix}
\nomenclature[a2]{$k$}{Time interval index}
\nomenclature[a3]{$L$}{Local-vertical/local-horizontal frame}
\nomenclature[a4]{$N_i$}{Multivariate normal distribution of dimension $i$}
\nomenclature[a5]{$N_p$}{Planning horizon}
\nomenclature[a6]{$n$}{Target orbit angular velocity}
\nomenclature[a7]{$n_L$}{LOS constraint grid size}
\nomenclature[a8]{$n_M$}{Reaction wheels constraint grid size}
\nomenclature[a9]{$n_T$}{Number of thrusters}
\nomenclature[b1]{$p$}{Thruster index}
\nomenclature[b2]{$\mathbf{R}$}{Rotation matrix}
\nomenclature[b3]{$r$}{Current MPC step}
\nomenclature[b4]{$\mathbf{r}_t$}{Target position with respect to Earth}
\nomenclature[b5]{$s$}{Standard deviation}
\nomenclature[b6]{$T$}{Interval duration}
\nomenclature[b7]{$t$}{Time}
\nomenclature[b8]{$\mathbf{u}$}{Velocity increment}
\nomenclature[b9]{$\mathbf{w}$}{Thruster pointing vector}
\nomenclature[c1]{$x$, $y$, $z$}{Relative position}
\nomenclature[c2]{$\pmb{\Theta}$}{Matrix full of zeros}
\nomenclature[c3]{$\theta_{rot}$}{Rotation angle}
\nomenclature[c4]{$\mu$}{Earth gravitational parameter}
\nomenclature[c5]{$\pmb{\sigma}$}{Modified Rodrigues parameter}
\nomenclature[c6]{$\pmb{\Phi}$}{Relative translational transition matrix}
\nomenclature[c7]{$\pmb{\omega}$}{Chaser angular velocity}
\nomenclature[c8]{$\| \cdot \|_l$}{$l$-norm of a vector}

\printnomenclature

\section{Introduction}

Autonomous spacecraft rendezvous and docking is becoming a more important topic in the space industry as access to space continues increasing. From the first rendezvous attempts (Gemini missions) to the Rosetta mission in 2014, the rendezvous manoeuvre has played a key role in different kinds of space missions such as Apollo, ISS, Hubble, etc. After decades of development, many approaches to achieve rendezvous for different mission profiles have been used, see \cite{Woffinden2007} for an historical review or \cite{Fehse2003} for the basics. Nowadays, an increasing interest to demonstrate autonomous rendezvous and flight formation operations for lightweight and low-power spacecraft is arising with CPOD, PRISMA and PROBA-3 missions as examples, see \cite{Roscoe2018, Persson2006, Castellani2013}.

Typically, the rendezvous problem has been widely studied just considering orbit control making the assumption that translational and rotational motions are decoupled. This problem has been usually tackled by means of direct transcription methods which transform the optimal control problem into a discrete optimization problem as in \cite{Lu2013, Breger2008, Richards2002, Gavilan2012, Vazquez2017} among others. The main advantage of these methods, against indirect ones, is that several kinds of constraints can be easily added to the problem such as approach corridors through the docking axis (V-bar or R-bar guidance), way-points, thrust direction inhibition, obstacle avoidance or fault-tolerant trajectories.

However, orbit and attitude control subsystems are mutually coupled, which is mainly due to the dependence of the thrusters orientation on the relative attitude between target and pursuer (at least in the short-term). Spacecraft attitude planning for direction reorientation manoeuvres, which are the ones needed to point the thrusters in an adequate way, is a topic with a vast literature. Reference \cite{Leve2015} proposed two dimensional attitude profiles with time derivatives saturation up to the jerk. Model Predictive Control (MPC) techniques based on linearization around a set point have been used in the works of \cite{Hegrenaes2005, Guiggiani2015}. A remarkable approach is the one followed by \cite{Louembet2009, Caubet2015} which is based on the attitude dynamics flatness property (see \cite{Fliess1995} for more details about flatness theory) that allowed them to transform the attitude dynamics into algebraic relations avoiding the need of numerical integration. 

Regarding previous works on six-degrees of freedom relative motion, adaptive tracking controllers based on feedback has been considered by \cite{Filipe2015} for rendezvous and by \cite{Wong2005, Wang2012} for flight-formation while strategies based on backstepping control have been employed by \cite{Kristiansen2008} for flight-formation and by \cite{Zhang2012_bis, Yan2016} for rendezvous operations. Sliding mode control has also been explored by \cite{Terui1998}. References \cite{Naasz2003, Siva2013, Moon2016, Wu2009} proposed a two stages approach, first they used an optimal control method for the translational motion, LQR in \cite{Naasz2003, Siva2013, Moon2016} or convex optimization in \cite{Wu2009}, and then they designed an attitude controller to obtain the orientations demanded by the translational plan. Reference \cite{Biggs2018} proposed a covering map based on quaternions to address 6-DOF open-loop motion planning based on basis functions and closed-loop kinematic feedback. The authors of the present paper proposed in \cite{Sanchez2018} a method based on the translational state transition matrix and the attitude flatness property to solve. Amongst the previous works,  dual quaternions, which contain information of both translational and rotational states, were used in \cite{Filipe2015, Wang2012}. Concerning the number of thrusters, \cite{Filipe2015, Kristiansen2008, Wang2012} assumed a pair of them available on each direction, six in total, whereas  \cite{Naasz2003} considered four thrusters and \cite{Siva2013, Wu2009, Moon2016, Yan2016, Sanchez2018} studied the case of single-thruster operations. The results of \cite{Biggs2018} are applied to both a classical six thrusters configuration and a single thruster one. The previous works assumed that torque is provided by an independent ACS system whereas \cite{Zhang2012_bis} considered six thrusters in a cuboid layout configuration providing both force and torque.
Apart from rendezvous and flight-formation operations, coupled motion has also been studied for geostationary satellites station-keeping \cite{Weiss2015} and solar sails control \cite{Gong2009}.  

In this paper, we consider a spacecraft equipped with reaction wheels and an arbitrary number of thrusters which seeks to rendezvous with a target flying in an eccentric orbit. The employed formulation allows to consider the coupled problem in an optimal way without any assumptions on the number of available thrusters which increase the applicability of the algorithm to different types of missions. In a similar way as \cite{Urbina2017}, a hybrid system is considered where the propulsive action is modelled as impulses but the attitude control is time continuous.

The proposed solution method transforms the time-continuous dynamics into algebraic relations by means of the translational state transition matrix and the attitude flatness property. Then, this equivalent optimal control problem is parameterized and discretized to obtain a finite tractable static program. Once an open-loop solution is obtained, a closed-loop MPC scheme, see \cite{Camacho2004}, based on linearization around the previously computed solution, is developed to reject disturbances and cope with unmodelled dynamics.   

The structure of this paper is as follows. Section \ref{model_rendezvous} describes the coupled translational and angular motion for spacecraft rendezvous. Next, Section \ref{planning_problem} presents the time-continuous rendezvous problem and its conversion to an equivalent problem. Section \ref{optimal_control} describes the employed methodology to solve this equivalent problem by means of parameterization and discretization. Section \ref{MPC_scheme} presents the linearized close-loop MPC scheme. Section \ref{results} shows results for cases of interest. Finally, Section \ref{conclusions} closes this paper with some additional considerations. 

\section{Model of Spacecraft Rendezvous}\label{model_rendezvous}

In this section, a six-degrees of freedom model for spacecraft rendezvous is presented. Firstly, the translational relative motion between the two vehicles is derived; secondly, the chaser angular motion model is described; and finally, both translational and angular motions are coupled.     

\subsection{Translational motion}\label{translational}

\begin{figure}[] 
\begin{center}
\includegraphics[width=10cm,height=10cm,keepaspectratio]{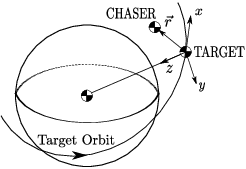}
\end{center}
\caption{LVLH frame}	
\label{LVLHframe}
\end{figure}

There is a considerable number of translational dynamic models for spacecraft rendezvous; the one to be chosen depends on the objectives and constraints on the mission. For instance, if the target vehicle is orbiting in a closed Keplerian orbit, the linearised equations of the relative position between an active chaser spacecraft and a passive target vehicle can be expressed in a cartesian reference frame as in \cite{Tschauner1965}, leading to the well known Tschauner-Hempel equations, or by means of its relative orbital elements as in \cite{Amico2006}. In this work, a cartesian reference frame is used
\begin{align}
\ddot{x}&= \ddot{\nu}z+2\dot{\nu}\dot{z}+\dot{\nu}^2x-\frac{\mu x}{r_t^3}+\sum^{n_T}_{p=1}\frac{F_{x,p}}{m}, \label{xHCW}\\
\ddot{y}&= -\frac{\mu y}{r_t^3}+\sum^{n_T}_{p=1}\frac{F_{y,p}}{m}, \label{yHCW}\\
\ddot{z}&= -\ddot{\nu}x-2\dot{\nu}\dot{x}+\dot{\nu}^2z+\frac{2\mu z}{r_t^3}+\sum^{n_T}_{p=1}\frac{F_{z,p}}{m}, \label{zHCW}
\end{align}
where $x$, $y$ and $z$ denote the position of the chaser in a local-vertical/local-horizontal (LVLH) frame of reference fixed on the center of gravity of the target vehicle (see Fig.\ref{LVLHframe}), in which $z$ refers to the radial position (positive pointing towards the centre of the Earth), $y$ to the cross-tack position (opposite to the orbit angular momentum) and $x$ closes the right-handed system (note that $x$ is not  necessarily aligned with the target velocity due to eccentricity). The velocity of the chaser in the LVLH frame is given by $\dot{x}$, $\dot{y}$ and $\dot{z}$; the variables $F_{x,p}$, $F_{y,p}$ and $F_{z,p}$ are the projections on the LVLH frame of the thrust force exerted by each one of the $n_T$ thrusters; and $m$ is the spacecraft mass which, for close enough rendezvous operations, is considered constant. The variables $r_t$ and $\nu$ are the target radius and true anomaly along its orbit, which are a function of time and its orbital elements (semi-major axis and eccentricity). The gravitation parameter of the Earth is $\mu$=398600.4 km$^3$/s$^2$.  

The state space model, $\mathbf{x}(t)$=$[x, \> y, \> z, \> \dot{x}, \> \dot{y}, \> \dot{z}]^T$, is governed by the linear time varying system (LTV) given by Eq.\eqref{xHCW}-\eqref{zHCW}. The independent variable of this LTV system can be changed from time to true anomaly leading to the simplified Tschauner-Hempel equations, see \cite{Tschauner1965}. A formal solution of the Tschauner-Hempel model by means of its state transition matrix, known as the Yamanaka-Ankersen matrix, was proposed by \cite{Yamanaka2002}. This transition matrix is computed by means of its fundamental matrix and inverse without need of numerical integration. In this work, following \cite{Vazquez2017} (note that the axes are not the same as in this work), the Yamanaka-Ankersen state transition matrix is expressed by means of the eccentric anomaly $E$,
\begin{equation}
\pmb{\Phi}(t,t_0)=\mathbf{Y}_{E(t)}\mathbf{Y}^{-1}_{E(t_0)}.\label{state_transition_matrix}
\end{equation}
Note that a one-to-one relation exist between time and eccentric anomaly through the Kepler equation
\begin{equation}
n(t-t_p)=E-e\sin E, \label{kepler_equation}
\end{equation}
where $t_p$ is the time at periapsis and is used as a reference point to measure $E$. The time $t_p$ is chosen such that it is equal or less than the starting manoeuvre time denoted by $t_0$ (substracting, if necessary, any number of orbital periods). Kepler's equation (\ref{kepler_equation}) is not analytically invertible, but its inverse can be found numerically with any desired degree of precision (see any Orbital Mechanics reference, such as \cite{Wie2008}).

Using the aforementioned state transition matrix, Eq.\eqref{state_transition_matrix}, and considering, as a simplification, an impulsive model, Eq.\eqref{impulsive_model}, the translational states transition equation is given by
\begin{equation}
\mathbf{x}(t)=\pmb{\Phi}(t,t_0)\mathbf{x}(t_0)+\sum^{k}_{i=0}\pmb{\Phi}(t,t_i)\mathbf{B}\mathbf{u}(t_i), \> \> \> \> t_{k} \leq t < t_{k+1}, \label{trans_state_prop}
\end{equation}   
where the input matrix is $\mathbf{B}$=$[\mathbf{\Theta}_{3\times3}, \> \mathbf{Id}]^T$, and the propulsive control signal $\mathbf{u}(t)$ is modelled as impulses (i.e. instantaneous changes of velocity) which describe with adequate accuracy a typical chemical thruster
\begin{equation}
\mathbf{u}(t)=\sum^{n_T}_{p=1}\sum^{N_p}_{k=0} \lim_{\Delta t_k \to 0} \int^{t_k+\Delta t_k}_{t_k} \frac{\mathbf{F}_p(t)}{m} dt=\sum^{n_T}_{p=1}\sum^{N_p}_{k=0}\Delta\mathbf{V}_{p}(t_k) \delta(t-t_k), \label{impulsive_model}
\end{equation}
being $\Delta \mathbf{V}_{p}$$\in$$\mathbb{R}^3$ the velocity increment given by the thruster $p$ and $N_p$+1$\in$$\mathbb{N}$ the number of thruster firings during the manoeuvre.

\subsection{Angular motion} 

In this section, the attitude representation parameter is chosen and some of their properties are presented. Then, the angular dynamics of a spacecraft considering only internal torques, which are the ones produced by reaction wheels, is derived. Finally, the attitude flatness property of the resulting angular dynamics is introduced (this property will be then exploited in Section \ref{planning_problem}). 

\subsubsection{Attitude representation and angular dynamics}

In this work, the modified Rodrigues parameters (MRP) representation (see \cite{Marandi1987, Schaub1996} for more details about MRP) is chosen rather than the widely used attitude quaternion. The modified Rodrigues parameters have the advantage of being a minimal attitude representation and are easier to linearize than attitude quaternions (incremental addition does not work for quaternions). Moreover, the unit-norm constraint of attitude quaternions is avoided in the problem formulation. The counterpart is that MRP suffer singularities when representing 3D rotations. The MRP are denoted as $\pmb{\sigma}$=$[\sigma_1, \> \sigma_2, \> \sigma_3]^T$ and its relation with the rotation angle, $\theta_{rot}$, and axis, $\mathbf{e}$, is
\begin{equation}
\pmb{\sigma}=\mathbf{e}\text{tan}(\theta_{rot}/4), \label{MRP_rot_param}
\end{equation}
where singularities arise when $\theta_{rot}$=$\pm2\pi$(2$j$-1)$\pi$ with $j$$\in$$\mathbb{N}$. However, they can be avoided by constraining $\theta_{rot}$$\in$$(-2\pi, \> 2\pi)$. The rotation matrix to change a vector from one reference frame to another is given by
\begin{equation}
\mathbf{R}(\pmb{\sigma}) = \mathbf{Id} + \frac{8 \pmb{\sigma}^{\times} \pmb{\sigma}^{\times} - 4(1 - \left\lVert \pmb{\sigma} \right\rVert_2^2)\pmb{\sigma}^{\times}}{\left(1 + \left\lVert \pmb{\sigma} \right\rVert_2^2 \right)^2}, \label{turn_matrix}
\end{equation}
where $\pmb{\sigma}^{\times}$$\in$$\mathbb{R}^{3\times3}$ is the cross product matrix, see \cite{Wie2008}. The attitude evolution of the chaser is defined by the kinematic and dynamic equations. The translational equations are expressed on a local frame so it is of interest to work with the attitude of the body frame with respect to the LVLH frame as in \cite{Wie2008}. The kinematics are given by
\begin{equation}
\dot{\pmb{\sigma}}(t) = \mathbf{C}(\pmb{\sigma}(t))[\pmb{\omega}(t)-\mathbf{R}(\pmb{\sigma}(t))\pmb{\omega}_{L/I}(t)], \label{att_kin}
\end{equation}   
being $\pmb{\omega}$=$[\omega_1, \> \omega_2, \> \omega_3]^T$ the angular velocity of the chaser body frame with respect to the inertial frame and $\pmb{\omega}_{L/I}$=$[0, \> -\dot{\nu}, \> 0]^T$ the angular velocity of the LVLH frame with respect to the inertial frame expressed on the local frame. The matrix $\mathbf{C}$ has the following expression
\begin{equation}
\mathbf{C}(\pmb{\sigma})=
\frac{1}{4}
\begin{bmatrix}
1 + \sigma_1^2 - \sigma_2^2 - \sigma_3^2 & 2(\sigma_1 \sigma_2 - \sigma_3) & 2(\sigma_1 \sigma_3 + \sigma_2)\\
2(\sigma_1 \sigma_2 + \sigma_3) & 1 - \sigma_1^2 + \sigma_2^2 - \sigma_3^2 & 2(\sigma_2 \sigma_3 - \sigma_1)\\
2(\sigma_1 \sigma_3 - \sigma_2) & 2(\sigma_2 \sigma_3 + \sigma_1) & 1 - \sigma_1^2 - \sigma_2^2 + \sigma_3^2\\
\end{bmatrix}.
\end{equation}
Additionally, the following equation describes the angular momentum variation, expressed on the chaser body frame, when the only considered torques are internal to the system (the ACS consists of reaction wheels)
\begin{equation}
\mathbf{I} \dot{\pmb{\omega}}(t) + \dot{\mathbf{H}}_{rw}(t) + \pmb{\omega}(t) \times \mathbf{H}_{tot} = \mathbf{0}, \label{euler_eq}
\end{equation}
where $\mathbf{I}$$\in$$\mathbb{R}^{3 \times 3}$ and $\mathbf{H}_{tot}$$\in$$\mathbb{R}^3$ are, respectively, the moment inertia matrix and the angular momentum of the spacecraft whereas $\mathbf{H}_{rw}$$\in$$\mathbb{R}^3$ is the angular momentum of the reaction wheels. Note that Eq.\eqref{att_kin}-\eqref{euler_eq} give the attitude evolution of the body frame, $B$, with respect to the inertial frame, $I$. From the fact that no external torques are applied, the spacecraft angular momentum is constant 
\begin{equation}
\mathbf{H}_{tot}=\mathbf{H}_b(t)+\mathbf{H}_{rw}(t)\equiv \text{constant} \label{kinetic_momentum_conservation},
\end{equation}
where $\mathbf{H}_b(t)$=$\mathbf{I}$$\pmb{\omega}(t)$ is the angular momentum of the platform. The attitude control signal is the exerted torque by the reaction wheels through its angular momentum variation, $\dot{\mathbf{H}}_{rw}(t)$.  

\subsubsection{Attitude flatness property} 

The angular motion given by Eq.\eqref{att_kin} and Eq.\eqref{euler_eq} is non-linear, hence accounting for them in the resolution of an optimal control problem usually require numerical integration, see \cite{Desai2008}. However, the considered angular dynamics has the flatness property and it is called a flat system, see \cite{Louembet2009}.

\textbf{Remark 1:} a flat system has a flat output which can be used to explicitly express all states and inputs in terms of the flat output and a finite number of its derivatives, see \cite{Fliess1995}.

Following \cite{Louembet2009}, the attitude representation parameter $\pmb{\sigma}(t)$ is chosen as the flat output. The differential equations of the angular motion, Eq.\eqref{att_kin} and Eq.\eqref{euler_eq}, can be transformed into algebraic relations, as a function of the flat output and its derivatives. Solving the angular velocity in Eq.\eqref{att_kin} and deriving the obtained expression with respect to time 
\begin{align}
\pmb{\omega}(t)&=\mathbf{C}^{-1}(\pmb{\sigma})\dot{\pmb{\sigma}}+\mathbf{R}(\pmb{\sigma})\pmb{\omega}_{L/I}, \label{flatness_omega}\\ 
\dot{\pmb{\omega}}(t)&=\mathbf{C}^{-1}(\pmb{\sigma})\left(\ddot{\pmb{\sigma}}+\dot{\mathbf{C}}(\pmb{\sigma})\mathbf{R}(\pmb{\sigma})\pmb{\omega}_{L/I}\right)+\dot{\mathbf{R}}(\pmb{\sigma})\pmb{\omega}_{L/I}+\mathbf{R}(\pmb{\sigma})\dot{\pmb{\omega}}_{L/I}-\dot{\mathbf{C}}(\pmb{\sigma})\pmb{\omega} \label{flatness_omega_dot},
\end{align}
and introducing  Eq.\eqref{flatness_omega}-\eqref{flatness_omega_dot} into Eq.\eqref{euler_eq}, the angular momentum variation of the reaction wheels is explicitly obtained as
\begin{equation}
\begin{array}{ll}
\dot{\mathbf{H}}_{rw}(t)=&-\mathbf{I}\left[\mathbf{C}^{-1}(\pmb{\sigma})\left(\ddot{\pmb{\sigma}}+\dot{\mathbf{C}}(\pmb{\sigma})\mathbf{R}(\pmb{\sigma})\pmb{\omega}_{L/I}\right)+\dot{\mathbf{R}}(\pmb{\sigma})\pmb{\omega}_{L/I}\right.\\
&\left.+\mathbf{R}(\pmb{\sigma})\dot{\pmb{\omega}}_{L/I}-\dot{\mathbf{C}}(\pmb{\sigma})\pmb{\omega}\right]-\left(\mathbf{C}^{-1}(\pmb{\sigma})\dot{\pmb{\sigma}}+\mathbf{R}(\pmb{\sigma})\pmb{\omega}_{L/I} \right) \times \mathbf{H}_{tot}. \label{flatness_kinetic_momentum_variation}
\end{array}
\end{equation}
Using the angular momentum conservation, Eq.\eqref{kinetic_momentum_conservation}, the reaction wheels angular momentum can also be expressed as a function of the flat output and its derivatives
\begin{equation}
\mathbf{H}_{rw}(t)=\mathbf{H}_{tot}-\mathbf{I}\mathbf{C}^{-1}(\pmb{\sigma})\dot{\pmb{\sigma}}, \label{flatness_kinetic_momentum}
\end{equation}
Note that time dependencies have been omitted at the right-hand side of Eq.\eqref{flatness_omega}-\eqref{flatness_kinetic_momentum} for clarity. 

\subsection{Coupling between translational and angular motion}

Now, the translational and angular motion coupling between the previous models is presented. The velocity increment given by each thruster $p$ on the LVLH frame, denoted by $L$, is
\begin{equation}
\Delta \mathbf{V}_p(t_k)=\mathbf{R}^T(\pmb{\sigma}(t_k)) \mathbf{w}_{p} u_p(t_k), \> \> \> \> u_p(t_k) \geq 0, \label{projected_increment_velocity}
\end{equation}
where $\mathbf{w}_{p}$$\in$$\mathbb{R}^3$ is a unit-norm vector representing the $p$ thruster orientation on the pursuer body frame and $u_{p}(t_k)$$\in$$\mathbb{R}$ is the impulse amplitude of the thruster $p$ at time $t_k$ and $\mathbf{R}(\pmb{\sigma})$$\in$$\mathbb{R}^{3\times3}$ is the rotation matrix between the chaser body frame and the LVLH frame. Introducing Eq.\eqref{projected_increment_velocity} into Eq.\eqref{impulsive_model}
\begin{equation}
\mathbf{u}(t)=\sum^{n_T}_{p=1}\sum^{N_p}_{k=0}\mathbf{R}^{T}(\pmb{\sigma}(t_k))\mathbf{w}_{p} u_p(t_k)\delta(t-t_k). \label{turnthrust}
\end{equation} 
 The coupling between translational and angular motion arises when the translational control input given by Eq.\eqref{turnthrust} is introduced into the translational states transition equation given by Eq.\eqref{trans_state_prop} leading to
\begin{equation}
\mathbf{x}(t)=\pmb{\Phi}(t,t_0)\mathbf{x}(t_0)+\sum^{k}_{i=0}\sum^{n_T}_{p=1}\pmb{\Phi}(t,t_i)\mathbf{B}\mathbf{R}^T(\pmb{\sigma}(t_i))\mathbf{w}_{p} u_p(t_i), \> \> \> \> t_{k} \leq t < t_{k+1}, \label{translational_transition_equation}
\end{equation}
Note that the propulsive action projected on the LVLH frame, $\mathbf{u}(t)$, depends on the vehicle attitude in a non-linear way by means of the rotation matrix between the pursuer body frame and the LVLH frame, see Eq.\eqref{turn_matrix}. The angular motion is not affected by the translational motion (gravity-gradient effects are neglected), hence Eq.\eqref{att_kin} and Eq.\eqref{euler_eq} still hold for the coupled model.

\section{Rendezvous planning problem} \label{planning_problem}

In this section, the objective function and constraints are presented. In a generic form, the rendezvous optimal control problem states as follows
\begin{equation}
\begin{aligned}
& \underset{u_{p}(t_k), \> \dot{\mathbf{H}}_{rw}(t)}{\text{minimize}}
& & J(u_p(t_k), \> \dot{\mathbf{H}}_{rw}(t)), \\
& \text{subject to}
& & \dot{\mathbf{v}}(t) = -2\pmb{\omega}\times\mathbf{v}-\dot{\pmb{\omega}}\times\mathbf{r}-\pmb{\omega}\times(\pmb{\omega}\times\mathbf{r})\\
&&& -\left(\mu/{\lVert \mathbf{r}_t \rVert}_2^3 \right)\left[\mathbf{r}-3\left(\mathbf{r}^T_t\mathbf{r}/{\lVert \mathbf{r}_t \rVert}_2^2\right)\mathbf{r}_t\right],\\
&&& \dot{\mathbf{r}}(t)=\mathbf{v}(t)+\sum^{N_p}_{k=0}\sum^{n_T}_{p=1}\mathbf{R}^T(\mathbf{\pmb{\sigma}}(t))\mathbf{w}_{p} u_p(t_k) \delta(t-t_k),\\
&&& \dot{\pmb{\omega}}(t) = -\mathbf{I}^{-1} \left[\dot{\mathbf{H}}_{rw}(t) + \pmb{\omega}(t) \times \mathbf{H}_{tot} \right],\\
&&& \dot{\pmb{\sigma}}(t) = \mathbf{C}(\pmb{\sigma}(t))[\pmb{\omega}(t)-\mathbf{R}(\pmb{\sigma}(t))\pmb{\omega}_{L/I}(t)],\\
&&& \mathbf{g}_1(\mathbf{r}(t)) \leq \mathbf{0}, \hspace{4.5cm} \text{LOS region}&\\
&&& \mathbf{g}_2(u_p(t_k)) \leq \mathbf{0}, \hspace{4.1cm} \Delta V \text{ bounds}&\\
&&& \mathbf{g}_3(\mathbf{H}_{rw}(t), \> \dot{\mathbf{H}}_{rw}(t)) \leq \mathbf{0}, \hspace{2.55cm} \text{ACS bounds}&\\
&&& \mathbf{g}_4(\mathbf{r}(t_0), \> \mathbf{v}(t_0), \> \pmb{\sigma}(t_0), \> \pmb{\omega}(t_0)) = \mathbf{0}, \hspace{1.2cm}  t_0 \text{ conditions}&\\
&&& \mathbf{g}_5(\mathbf{r}(t_f), \> \mathbf{v}(t_f), \> \pmb{\sigma}(t_f), \> \pmb{\omega}(t_f)) = \mathbf{0}, \hspace{1.05cm} t_f \text{ conditions}&
\end{aligned}\label{opt_orig}
\end{equation}
where $\mathbf{r}(t)$=$[x(t), \> y(t), \> z(t)]^T$, $\mathbf{v}(t)$=$[\dot{x}(t), \> \dot{y}(t), \> \dot{z}(t)]^T$ and $\mathbf{r}_t(t)$=$[0, \> 0, \> -r_t(t)]^T$. Note that time dependencies have been omitted at the right hand side of the translational dynamics equation for clarity. The control inputs are the thrusters impulses amplitudes at the firing times (which are known beforehand) and the reaction wheels angular momentum variation. Next the objective function and constraints appearing in \eqref{opt_orig} are detailed in \eqref{obj_function} and \eqref{constraints} respectively.

\subsection{Objective function} \label{obj_function}

The chosen objective function seeks to minimize fuel consumption, which is equivalent to minimize the $L^1$-norm of the applied impulses
\begin{equation}
J=\sum^{N_p}_{k=0}\sum^{n_T}_{p=1} u_{p}(t_k). \label{obj_func}  
\end{equation}
Note that the absolute value symbol is not needed since $u_{p}(t_k)$ are always positive because of Eq.\eqref{projected_increment_velocity}. Moreover, reaction wheels use electrical power and therefore their associated cost do not appear in Eq.\eqref{obj_func}.

\subsection{Constraints of the problem} \label{constraints}

Three sets of constraints are considered in this paper. Firstly, path constraints on the relative translational states ($\mathbf{g}_1$); secondly, the control variables (impulses amplitudes and reaction wheels angular momentum) are bounded ($\mathbf{g}_2$ and $\mathbf{g}_3$); and finally, initial and terminal states values are prescribed ($\mathbf{g}_4$ and $\mathbf{g}_5$).

\subsubsection{Path constraints}

\begin{figure}[] 
\begin{center}
\includegraphics[width=8cm,height=8cm,keepaspectratio]{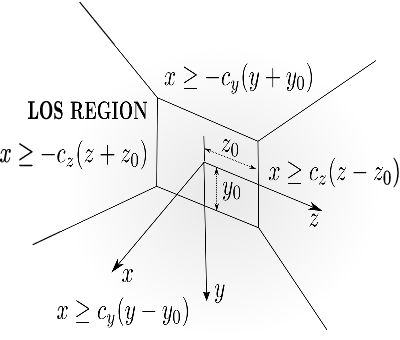}
\end{center}
\caption{LOS region}
\label{LOSregion}
\end{figure}

For sensing purposes (see \cite{Breger2008}), it is required that the chaser vehicle remains inside a line of sight (LOS) area from the docking port, thus guaranteeing that the chaser spacecraft is at all time visible from the docking port. The LOS region can be defined by the equations $x$$\geq$$c_y(y-y_0)$, $x$$\geq $$-c_y(y+y_0)$, $x$$\geq$$c_z(z-z_0)$, $x$$\geq$$-c_z(z+z_0)$ and $x$$\geq$0; these equations limit the relative translational state space by five planes as shown in Fig.\ref{LOSregion}.
One can define the LOS constraint algebraically, at any instant $t$, as $\mathbf{A}_L \mathbf{x}(t) \leq \mathbf{b}_L$, where
\begin{equation}
\mathbf{A}_L=
\begin{bmatrix}
-1 & c_y & 0 & 0 & 0 & 0\\
-1 & -c_y & 0 & 0 & 0 & 0\\
-1 & 0 & c_z & 0 & 0 & 0\\
-1 & 0 & -c_z & 0 & 0 & 0\\ 
-1 & 0 & 0 & 0 & 0 & 0\\
\end{bmatrix}, \> \> \>
\mathbf{b}_L=
\begin{bmatrix}
c_y y_0\\
c_y y_0\\
c_z z_0\\
c_z z_0\\
0\\ 
\end{bmatrix}.\label{LOS_eq}
\end{equation}

\subsubsection{Control bounds}

Regarding the thrusters performance, it is assumed that the impulse amplitude provided by each thruster is bounded above (and below by zero)
\begin{equation}
0 \leq u_{p}(t_k) \leq u_{p,max}, \> \> \> \> p=1 \hdots n_T. \label{thrusters_constraints}
\end{equation}  
Note that $u_{p}$ can take any value in the allowed interval (it is assumed that thrusters opening times can be adjusted to produce the exact impulse amount).

On the other hand, each one of the reaction wheels saturates when it stores a certain amount of angular momentum which is equivalent to consider that reaction wheels velocities are limited. Moreover limits on the angular acceleration exist for each wheel
\begin{align}
-H_{i,max} \leq H_{i,rw}(t) \leq H_{i,max}, \> \> \> \> i=1,2,3, \label{angular_momentum_constraint} \\
-\dot{H}_{i,max} \leq \dot{H}_{i,rw}(t) \leq \dot{H}_{i,max}, \> \> \> \> i=1,2,3. \label{angular_momentum_variation_constraint}
\end{align}

\subsubsection{Boundary constraints}

The chaser is assumed to depart from a given point and velocity with a given orientation and angular velocity
\begin{equation}
\mathbf{x}(t_0)=\mathbf{x}_0, \> \> \> \> \pmb{\sigma}(t_0) = \pmb{\sigma}_0, \> \> \> \> \pmb{\omega}(t_0)=\pmb{\omega}_0,
\end{equation}  
and it has to met prescribed states at the end of the manoeuvre
\begin{equation}
\mathbf{x}(t_f)=\mathbf{x}_f, \> \> \> \> \pmb{\sigma}(t_f) = \pmb{\sigma}_f, \> \> \> \> \pmb{\omega}(t_f)=\pmb{\omega}_f,
\end{equation}  
where the last three components of $\mathbf{x}_f$ are null and $\pmb{\omega}_f$=$\mathbf{R}(\pmb{\sigma}_f)$$\pmb{\omega}_{L/I}(t_f)$ to have no relative angular velocity between the body and LVLH frame.

\subsection{Equivalent rendezvous planning problem}

The aim of the optimal control problem \eqref{opt_orig} is to guarantee rendezvous with the target along a prescribed approach region (LOS) while respecting control bounds and minimizing fuel consumption. Using the coupled transition equation for the translational states, see Eq.\eqref{translational_transition_equation}, and the algebraic relations derived from the attitude flatness property, developed through  Eq.\eqref{flatness_omega}-\eqref{flatness_kinetic_momentum}, we formulate an equivalent planning problem, where differential equations are replaced by algebraic ones (as a function of the flat output and its derivatives), without losing any information, 
\begin{equation}
\begin{aligned}
& \underset{u_{p}(t_k), \> \pmb{\sigma}(t)}{\text{minimize}}
& & \sum^{N_p}_{k=0}\sum^{n_T}_{p=1} u_{p}(t_k), \\
& \text{subject to}
& & \mathbf{x}(t)=\pmb{\Phi}(t,t_0)\mathbf{x}(t_0)+\sum^{k}_{i=0}\sum^{n_T}_{p=1}\pmb{\Phi}(t,t_i)\mathbf{B}\mathbf{R}^T(\pmb{\sigma}(t_i))\mathbf{w}_{p} u_p(t_i),\\
&&& \> \> \> \> t_{k} \leq t < t_{k+1},\\
&&& \mathbf{A}_L \mathbf{x}(t) \leq \mathbf{b}_L,\\
&&& 0 \leq u_p(t) \leq u_{p,max}, \> \> \> \> p=1 \hdots n_T,\\
&&& -H_{i,max} \leq H_{i,rw}(\pmb{\sigma}(t), \dot{\pmb{\sigma}}(t)) \leq H_{i,max}, \> \> \> \> i=1,2,3,\\
&&& -\dot{H}_{i,max} \leq \dot{H}_{i,rw}(\pmb{\sigma}(t), \dot{\pmb{\sigma}}(t), \ddot{\pmb{\sigma}}(t)) \leq \dot{H}_{i,max}, \> \> \> \> i=1,2,3,\\
&&& \mathbf{x}(t_0)=\mathbf{x}_0, \> \> \> \> \pmb{\sigma}(t_0)=\pmb{\sigma}_0, \> \> \> \> \pmb{\omega}(t_0)=\pmb{\omega}_0,\\
&&& \mathbf{x}(t_f)=\mathbf{x}_f, \> \> \> \> \pmb{\sigma}(t_f)=\pmb{\sigma}_f, \> \> \> \> \pmb{\omega}(t_f)=\pmb{\omega}_f.\\
\end{aligned}\label{opt_equiv}
\end{equation}
Beside the fact that the equivalent planning problem \eqref{opt_equiv} is integration free, it is still infinite dimensional. In the next section, it is shown how to make this problem tractable by means of parameterization and discretization.

\section{Optimal control computation}\label{optimal_control}

In this section, the resolution method to the equivalent optimal control problem \eqref{opt_equiv} is presented. The proposed methodology is based on a B-spline parameterization of the flat output (MRP) and the discretization of the infinite dimensional constraints. The result is a tractable NLP problem which needs an initial guess to be solved.    

\subsection{Non-linear programming description}

First of all, it is considered that the $N_p$+1 impulses application times are equally spaced through the manoeuvre time, $t$$\in$$[t_0, \> t_f]$, with timespan $T$=$(t_f-t_0)/N_p$, hence $t_k$=$t_0$+$kT$ for $k$=$0 \hdots N_p$. These firing times will be denoted as nodes.

\subsubsection{B-splines parameterization of the flat output}

The attitude flatness property allows any kind of MRP time evolution parameterization. In this work, following \cite{Louembet2009}, B-splines, see \cite{Kress1998} for more details about them, are chosen to parameterize the flat output since they define flexible trajectories with a high degree of differentiability using a low number of parameters
\begin{equation}
\pmb{\sigma}(t)=\sum^{n_c}_{j=1}\mathbf{a}_{j}B_{j,q}(t), \label{MRP_parameterised}
\end{equation}
where the $B_{j,q}(t)$ are $q$th order B-splines built on the knots sequence, $\mathbf{t}_{knots}\in\mathbb{R}^{n_{knots}}$, and the $\mathbf{a}_{j}$$\in$$\mathbb{R}^3$ coefficients are called control points.

\textbf{Remark 2:} The B-splines intrinsically assure continuity up to $C^{q}$. Given the order $q$ and the number of coefficients $n_c$, the number of knots must satisfy $n_{knots}$=$n_c$+$q$+1. 

The attitude profile has to be continuous up to its second derivative, hence, $q$$\geq$2. On the other hand, it is chosen to have at least one control point to represent the attitude at each node plus four additional control points to impose $\dot{\pmb{\sigma}}$ and $\ddot{\pmb{\sigma}}$ at the beginning and end of the manoeuvre. The previous consideration leads to $n_c$=$N_p$+5, therefore $n_{knots}$=$N_p$+$q$+6. The knots are selected as the nodes $t_k$, augmented at left and right by repeating $t_0$ and $t_f$  
\begin{equation}
\mathbf{t}_{knots}=[t_0, \hdots, t_0, \> t_1, \hdots, t_{N_p-1}, \> t_{f}, \hdots, t_{f}]^T. \label{knots_placement}
\end{equation}     

\subsubsection{Discretization of time continuous constraints}

The time continuous constraints are the path constraint related to the LOS region, see Eq.\eqref{LOS_eq}, and the bounds on the reaction wheels angular momentum and its variation, Eq.\eqref{angular_momentum_constraint}-\eqref{angular_momentum_variation_constraint}. Each one of these constraints is discretized with a time grid within each interval, $k$. The LOS constraint is gridded with $n_L$ equally spaced subintervals of duration $T_L$=$T/n_L$ at which the constraint is imposed
\begin{equation}
\mathbf{A}_L \mathbf{x}(t_{k,l}) \leq \mathbf{b}_L,  \> \> \> \> 
t_{k,l}=t_0+(k-1)T+lT_L, \> \> \> \> k=1 \hdots N_p, \> \> \> \> l=1 \hdots n_L,
\end{equation}
whereas the reaction wheels constraints are gridded with $n_M$ equally spaced subintervals of duration $T_M$=$T/n_M$
\begin{equation}
\begin{aligned}
-H_{i,max} \leq H_i (t_{k,m}, \mathbf{a}_j) \leq H_{i,max}, \> \> \> \>& i=1,2,3,\\
-\dot{H}_{i,max} \leq \dot{H}_i (t_{k,m}, \mathbf{a}_j) \leq \dot{H}_{i,max}, \> \> \> \>& i=1,2,3,\\
t_{k,m}=t_0+(k-1)T+mT_M, \> \> \> \>& m=0 \hdots n_{M}.
\end{aligned}
\end{equation}

\subsubsection{Discrete optimization problem}

To ease the notation, following \cite{Gavilan2012}, a compact formulation of the discrete problem is developed. Defining the following stack vectors $\mathbf{x_S}$$\in$$\mathbb{R}^{6n_LN_p}$, $\mathbf{u}_{\mathbf{S}p}\in\mathbb{R}^{N_p+1}$ and $\mathbf{a_S}$$\in$$\mathbb{R}^{3n_c}$ as  
\begin{align}
\mathbf{x_S}&=[\mathbf{x}_{1,1}^T, \hdots, \> \mathbf{x}_{1,n_L}^T, \> \mathbf{x}_{2,1}^T, \hdots, \> \mathbf{x}_{2,n_L}^T, \> \mathbf{x}_{3,1}^T, \hdots \hdots, \> \mathbf{x}_{N_p,n_L}^T]^T, \label{init}\\ \
\mathbf{u}_{\mathbf{S}p}&=[u_{p,0}, \> u_{p,1}, \hdots, \> u_{p,N_p}]^T,\\
\mathbf{a_S}&=[\mathbf{a}_1^T, \> \mathbf{a}_2^T, \hdots, \> \mathbf{a}_{n_c}^T]^T,
\end{align}
and the stack matrices $\mathbf{F}$$\in$$\mathbb{R}^{6n_LN_p \times 6}$ and $\mathbf{G}_p$$\in$$\mathbb{R}^{6n_LN_p \times (N_p+1)}$ 
\begin{equation}
\mathbf{F}=[\pmb{\Phi}^T(t_{1,1},t_0), \hdots, \pmb{\Phi}^T(t_{1,n_L},t_0), \pmb{\Phi}^T(t_{2,1},t_0), \hdots \hdots, \pmb{\Phi}^T(t_{N_p,n_L},t_0)]^T,
\end{equation}
\begin{equation}
\begin{aligned}
&\mathbf{G}_p=\\
&\begin{bmatrix}
\pmb{\Phi}(t_{1,1}, t_{0})\mathbf{B}\mathbf{R}^{T}(\pmb{\sigma}_0)\mathbf{w}^{B}_{p} & \mathbf{\Theta}_{6\times1} & \hdots & \mathbf{\Theta}_{6\times1}\\
\vdots & \vdots & \ddots & \vdots \\
\pmb{\Phi}(t_{1,n_L}, t_{0})\mathbf{B}\mathbf{R}^{T}(\pmb{\sigma}_0)\mathbf{w}^{B}_{p} & \mathbf{\Theta}_{6\times1} & \hdots & \mathbf{\Theta}_{6\times1}\\
\pmb{\Phi}(t_{2,1}, t_{0})\mathbf{B}\mathbf{R}^{T}(\pmb{\sigma}_0)\mathbf{w}_{p} & \pmb{\Phi}(t_{2,1}, t_{1})\mathbf{B}\mathbf{R}^{T}(\pmb{\sigma}_1)\mathbf{w}_{p} & \hdots & \mathbf{\Theta}_{6\times1}\\
\vdots & \vdots & \ddots & \vdots\\
\pmb{\Phi}(t_{2,n_L}, t_{0})\mathbf{B}\mathbf{R}^{T}(\pmb{\sigma}_0)\mathbf{w}_{p} & \pmb{\Phi}(t_{2,n_L}, t_{1})\mathbf{B}\mathbf{R}^{T}(\pmb{\sigma}_1)\mathbf{w}_{p} & \hdots & \mathbf{\Theta}_{6\times1}\\
\pmb{\Phi}(t_{3,1}, t_{0})\mathbf{B}\mathbf{R}^{T}(\pmb{\sigma}_0)\mathbf{w}_{p} & \pmb{\Phi}(t_{3,n_L}, t_1)\mathbf{B}\mathbf{R}^{T}(\pmb{\sigma}_1)\mathbf{w}_{p} & \hdots & \mathbf{\Theta}_{6\times1}\\
\vdots & \vdots & \ddots & \vdots\\
\vdots & \vdots & \ddots & \vdots\\
\pmb{\Phi}(t_{N_p,n_L}, t_{0})\mathbf{B}\mathbf{R}^{T}(\pmb{\sigma}_0)\mathbf{w}_{p} & \pmb{\Phi}(t_{N_p,n_L}, t_{1})\mathbf{B}\mathbf{R}^{T}(\pmb{\sigma}_1)\mathbf{w}_{p} & \hdots & \mathbf{B}\mathbf{R}^{T}(\pmb{\sigma}_{N_p})\mathbf{w}_{p}\\
\end{bmatrix}, \label{fin}
\end{aligned}
\end{equation} 
where $\pmb{\sigma}_k$=$\pmb{\sigma}(t_k,\mathbf{a_S})$ represents the attitude at each node. The relation between the stack vectors and matrices defined in Eq.(\ref{init})-(\ref{fin}) is given by
\begin{equation}
\mathbf{x_S} = \mathbf{F} \mathbf{x}_0 + \sum^{n_T}_{p=1}\mathbf{G}_p(\mathbf{a_S})\mathbf{u}_{\mathbf{S}p}. \label{compact1}
\end{equation}
Now, the infinite dimensional problem \eqref{opt_equiv} boils down to NLP, expressed with the compact formulation, by means of the above parameterization and continuous constraints discretization
\begin{equation}
\begin{aligned}
& \underset{\mathbf{u}_{\mathbf{S}p}, \> \mathbf{a_S}}{\text{minimize}}
& & \sum^{n_T}_{p=1}{\lVert \mathbf{u}_{\mathbf{S}p}\rVert}_1, \\
& \text{subject to}
& & \mathbf{A}_{LS}\sum^{n_T}_{p=1}\mathbf{G}_p(\mathbf{a_S})\mathbf{u}_{\mathbf{S}p} \leq \mathbf{b}_{LS}-\mathbf{A}_{LS}\mathbf{F}\mathbf{x}_0, \\
&&& \mathbf{0} \leq \mathbf{u}_{\mathbf{S}p} \leq \mathbf{u}_{\mathbf{S}p,max}, \> \> \> \> p=1\hdots n_T, \\
&&& -H_{i,max} \leq H_i(t_{k,m}, \> \mathbf{a_S}) \leq H_{i,max}, \> \> i=1,2,3, \\
&&& -\dot{H}_{i,max} \leq \dot{H}_i(t_{k,m}, \> \mathbf{a_S}) \leq \dot{H}_{i,max}, \> \> i=1,2,3, \\
&&& \mathbf{A}_{rend}\sum^{n_T}_{p=1}\mathbf{G}_p(\mathbf{a_S})\mathbf{u}_{\mathbf{S}p} = \mathbf{x}_f-\mathbf{A}_{rend}\mathbf{F}\mathbf{x}_0, \\
&&& \pmb{\sigma}(t_0, \> \mathbf{a_S}) = \pmb{\sigma}_0, \> \> \dot{\pmb{\sigma}}(t_0, \> \mathbf{a_S}) = \mathbf{0}, \> \> \ddot{\pmb{\sigma}}(t_0, \> \mathbf{a_S}) = \mathbf{0},\\
&&& \pmb{\sigma}(t_f, \> \mathbf{a_S}) = \pmb{\sigma}_f, \> \> \dot{\pmb{\sigma}}(t_f, \> \mathbf{a_S}) = \mathbf{0}, \> \> \ddot{\pmb{\sigma}}(t_f, \> \mathbf{a_S}) = \mathbf{0},
\end{aligned}\label{opt_NLP}
\end{equation}
where $\mathbf{A}_{LS}$$\in$$\mathbb{R}^{5n_LN_p \times 6n_LN_p}$ and $\mathbf{b}_{LS}$$\in$$ \mathbb{R}^{5n_LN_p}$ stack the LOS matrix (diagonally) and vector, see Eq.(\ref{LOS_eq}), respectively. The parameters $\mathbf{u}_{\mathbf{S}p,max}$$\in$$\mathbb{R}^{N_p+1}$ are stack vectors whose components are all equal to $u_{p,max}$. The matrix $\mathbf{A}_{rend}=[\mathbf{\Theta}_{6\times6(n_LN_p-1)}, \> \mathbf{Id}_{6\times6}]$ is employed to impose the rendezvous condition. It has been considered that reaction wheels kinetic momentum variation at initial and final time shall be zero which constrains $\ddot{\pmb{\sigma}}$. A NLP solver is required to obtain a solution of the static program \eqref{opt_NLP}.

\subsection{Initial guess computation (hotstart)} \label{initial_guess}

Any NLP solver needs an initial guess to compute the optimal solution of problem  \eqref{opt_NLP}. In this case, the process is composed of two steps; first, a traditional six-thrusters spacecraft model with three-degrees of freedom is employed to formulate and solve a linear programming (LP) problem; and then, this obtained LP solution is converted to NLP decision variables, $\mathbf{u}_{\mathbf{S}p}$ and $\mathbf{a_S}$.

\subsubsection{Six-thrusters problem formulation}

Considering a pair of thrusters available for each LVLH axis, the control can be expressed at each node as $\mathbf{u}_k$=$[\Delta V_{x,k}, \> \Delta V_{y,k}, \> \Delta V_{z,k}]^T$, hence, the translational states transition equation is linear
\begin{equation}
\mathbf{x}(t)=\pmb{\Phi}(t, \> t_0)\mathbf{x}_0+\sum^{k}_{i=0}\pmb{\Phi}(t, \> t_i)\mathbf{B}\mathbf{u}_i, \> \> \> \> t_{k} \leq t < t_{k+1},
\end{equation}
and the LP problem is posed as
\begin{equation}
\begin{aligned}
& \underset{\mathbf{u}_k}{\text{minimize}}
& & \sum^{N_p}_{k=0}\| \mathbf{u}_k \|_1, \\
& \text{subject to}
& & \mathbf{A}_{L}\mathbf{x}_(t_{k,l}) \leq \mathbf{b}_{L}, \\
&&& -\text{max}(u_{p,max})/\sqrt{3} \leq \mathbf{u}_k \leq \text{max}(u_{p,max})/\sqrt{3}, \\
&&& \mathbf{x}_{N_p}=\mathbf{x}_f,\\
&&& \mathbf{A}_{\mathbf{u}_{0}}\mathbf{u}_{0}=\mathbf{0},\\ &&&\mathbf{A}_{\mathbf{u}_{N_p}}\mathbf{u}_{N_p}=\mathbf{0},\\
\end{aligned}\label{opt_LP}
\end{equation}
where the bounds on the impulse amplitude for each direction have been conservatively chosen to not overpass the upper bound of the thruster with more available impulse amplitude when all thrusters saturate (thus the use of $\sqrt{3}$). The purpose of the last linear constraints, expressed by means of the matrices $\mathbf{A}_{\mathbf{u}_{0}}$ and $\mathbf{A}_{\mathbf{u}_{N_p}}$, is to make the initial and final orientations compatibles with the initial, $\pmb{\sigma}_0$, and desired final attitude, $\pmb{\sigma}_f$, respectively. It should be noted that the $L^1$-norm term in the objective function is non-linear because $\mathbf{u}_k$ can take both positive or negative values. However, this issue is avoided by adding optimization slack variables only allowed to take positive values. 

\subsubsection{Six-thrusters solution transformation to a NLP solution}

Once the LP problem \eqref{opt_LP} is solved, the impulses amplitudes on each thruster are chosen as $u_{1,k}$=$\|\mathbf{u}_k\|_2$ and $u_{p \neq 1,k}$=$0$. The thruster labelled with $p$=1 is the one with higher $u_{p,max}$.
 
The B-spline control points, $\mathbf{a_S}$, are obtained matching the $N_p$+1 demanded orientations at the nodes by the LP solution. The MRP at the nodes can be obtained with the aid of the rotation angle and axis. First, denote by $k^{*}_i$, where the subscript $i$ refers to the number of required thruster firings ($\|\mathbf{u}_{k^{*}_i}\|_2>$0), the nodes at which a non-null impulse amplitude is demanded or an attitude has to be reached (instant $t_{N_p}$) and then compute the unitary vector $\mathbf{z}_{k^*_i}$ representing the velocity increment orientation, expressed on the inertial frame since attitude is defined between the chaser body frame and the inertial frame, at these nodes
\begin{equation}
\mathbf{z}_{k^*_i} = [u_{x,k}, u_{y,k}, u_{z,k}]^T/\|\mathbf{u}_k\|_2, \> \> \> \> k^{*}_i=k, \> \> \> \> \text{if} \> \> \|\mathbf{u}_k\|_2>0.\label{impulse_orientation}
\end{equation}
Using $\mathbf{z}_{k^*_i}$, it is possible to obtain the rotation MRP, $\pmb{\sigma}_{rot}$, between consecutive orientations. For the nodes without thruster firings ($\|\mathbf{u}_k\|_2$=$0$), the attitude at this node $k$ is chosen as the value of the interpolated MRP, between the nodes $k^*_{i-1}$ and $k^*_i$, evaluated at the instant $t_k$. The rotation MRP between $t_{k-1}$ and $t_k$ is
\begin{equation}
\pmb{\sigma}_{rot_{k/k-1}}=\mathbf{e}_{k^*_i}\text{tan}(s_k \theta_{k^*_i}/4),\> \> \> \> s_k = \frac{k-k^*_{i-1}}{k^*_i-k^*_{i-1}},\> \> \> \> t_{k-1}, t_k \in [t_{k^*_{i-1}}, \> t_{k^*_i}]. \label{inter_attitude}
\end{equation}
where the rotation angle and axis of Eq.\eqref{inter_attitude} are obtained using the previously computed orientations, see Eq.\eqref{impulse_orientation} 
\begin{align}
\theta_{k^*_i}&=\textup{acos}(\mathbf{z}_{k^*_i} \cdot \mathbf{z}_{k^*_{i-1}}), \label{eulerangle}\\
\mathbf{e}_{k^*_i}&=\frac{\mathbf{z}_{k^*_i} \times \mathbf{z}_{k^*_{i-1}}}{\lVert\mathbf{z}_{k^*_i} \times \mathbf{z}_{k^*_{i-1}}\rVert_2}.\label{euleraxis}
\end{align}
This way, smooth attitude transitions are obtained when some nodes do not have burnings. 
Since $\theta_{m_i}$$\in$$[-\pi, \> \pi]$, no singularities arise when computing $\pmb{\sigma}_{rot}$. Once the rotation MRP is obtained, it is possible to compute the MRP at each node $t_k$. The MRP at the nodes are determined applying the attitude composition rule given by
\begin{equation}
\begin{aligned}
&\pmb{\sigma}_k =\\
&\frac{(1-\lVert \pmb{\sigma}_{rot_{k/k-1}} \rVert^2_2)\pmb{\sigma}_{k-1}+(1-\lVert \pmb{\sigma}_{k-1} \rVert^2_2)\pmb{\sigma}_{rot_{k/k-1}}+2\pmb{\sigma}_{k-1}\times\pmb{\sigma}_{rot_{k/k-1}}}{1+(\lVert \pmb{\sigma}_{rot_{k/k-1}} \rVert_2 \lVert \pmb{\sigma}_{k-1} \rVert_2)^2 - 2\pmb{\sigma}_{rot_{k/k-1}}\cdot\pmb{\sigma}_{k-1}}. \label{att_rot}
\end{aligned}
\end{equation}
The last step is to compute the control points for this nodes sequence. Imposing null $\dot{\pmb{\sigma}}$ and $\ddot{\pmb{\sigma}}$ at $t_0$ and $t_f$ a linear system of $3n_c$ equations with $3n_c$ unknowns (remember that $n_c$=$N_p$+5) can be easily solved to obtain the initial guess B-splines control points $\mathbf{a_S}$ as it is proposed in \cite{Kress1998}. 

\section{MPC scheme} \label{MPC_scheme}

Once the NLP problem \eqref{opt_NLP} is solved, an open-loop solution for the rendezvous manoeuvre is available. However, disturbances, unmodelled dynamics, etc., will perturb the planned path while the spacecraft is manoeuvring, hence a MPC scheme, based on linearization around this previously computed solution, is developed in this section. The trajectory is recomputed on-line, in a sliding horizon framework, by solving a quadratic programming problem after each sampling interval which eases the computational burden (compared to the NLP) and does not need an initial guess. The terminal constraints are relaxed, considering them as terminal costs instead of constraints, to prevent feasibility issues and augment stability. 

\subsection{Linearized model}

Allowing small increments of the decision variables, $\Delta u_{p,k}$ and $\Delta \mathbf{a}_j$, the translational transition Eq.\eqref{translational_transition_equation} states as follows
\begin{equation}
\begin{aligned}
\mathbf{x}(t)=&\pmb{\Phi}(t, t_r)\mathbf{x}_{r}+\sum^{r+k}_{i=r}\sum^{n_T}_{p=1}\pmb{\Phi}(t,t_i)\mathbf{B}\bigg(\mathbf{R}^T(\pmb{\sigma}(t_i))\mathbf{w}_p(u_{p,i}+\Delta u_{p,i})\\
&+ \sum^{r+n_c}_{j=r+1} \Delta\mathbf{R}_{p,\mathbf{a}_j}(\pmb{\sigma}(t_i))u_{p,i} \Delta \mathbf{a}_j \bigg), \> \> t \in [t_{r+k},t_{r+k+1}), \label{linearized_translational_transition_equation}
\end{aligned}
\end{equation}
where $r$=$1 \hdots N_p$ is the current MPC step and the matrix $\Delta\mathbf{R}_{p,\mathbf{a}_j}$$\in$$\mathbb{R}^{3\times 3}$ has the following expression
\begin{equation}
\Delta R_{(p,\mathbf{a}_j),\alpha \beta}=\sum_{\beta=1}^{3} \left.\frac{\partial{\left(R^T_{\alpha \beta}w_{p,\beta}\right)}}{\partial{\sigma_\beta}}\frac{\partial \sigma_\beta}{\partial{a_{j,\beta}}}\right|_{\pmb{\sigma}_{i}, \mathbf{a}_j}, \> \> \> \> \alpha=1,2,3, \> \> \> \> \beta=1,2,3.
\end{equation}
Following with the compact formulation developed through Eq.\eqref{init}-\eqref{compact1}, one can define the following stack vectors $\Delta \mathbf{u}_{\mathbf{S}p}(r)$$\in$$\mathbb{R}^{N_p+1}$, $\Delta \mathbf{a_S}(r)$$\in$$\mathbb{R}^{3n_c}$
\begin{align}
\Delta \mathbf{u}_{\mathbf{S}p}(r)&=[\Delta u_{p,r}, \> \Delta u_{p,r+1}, \hdots, \> \Delta u_{p,r+N_p}]^T,\\
\Delta \mathbf{a_S}(r)&=[\Delta \mathbf{a}_{r+1}^T, \> \Delta \mathbf{a}_{r+2}^T, \hdots, \> \Delta \mathbf{a}_{r+n_c}^T]^T,
\end{align}
and the stack matrix $\mathbf{H}_p(r)$$\in$$\mathbb{R}^{6n_LN_p \times 3n_c}$
\begin{equation*}
\begin{aligned}
&\mathbf{H}_p(r)=\\
&\begin{bmatrix}
\pmb{\Phi}(t_{r+1,1}, t_{r})\mathbf{B}\Delta\mathbf{R}_{p,\mathbf{a}_{r+1}} u_{p,r} &  \hdots & \pmb{\Phi}(t_{r+1,1}, t_{r})\mathbf{B}\Delta\mathbf{R}_{p,\mathbf{a}_{r+n_c}} u_{p,r}\\
\vdots & \ddots & \vdots \\
\pmb{\Phi}(t_{r+1,n_L}, t_{r})\mathbf{B}\Delta\mathbf{R}_{p,\mathbf{a}_{r+1}} u_{p,r} & \hdots & \pmb{\Phi}(t_{r+1,n_L}, t_{r})\mathbf{B}\Delta\mathbf{R}_{p,\mathbf{a}_{r+n_c}} u_{p,r}\\
\vdots & \ddots & \vdots \\
\vdots & \ddots & \vdots \\
\sum\limits^{r+N_p}_{i=r}\pmb{\Phi}(t_{r+N_p,n_L}, t_{i})\mathbf{B}\Delta\mathbf{R}_{p,\mathbf{a}_{r+1}} u_{p,i} & \hdots & \sum\limits^{r+N_p}_{i=r}\pmb{\Phi}(t_{r+N_p,n_L}, t_{i})\mathbf{B}\Delta\mathbf{R}_{p,\mathbf{a}_{r+n_c}} u_{p,i}\\
\end{bmatrix}, \label{compact_H}
\end{aligned}
\end{equation*}  
where one should note that many of the matrices $\Delta\mathbf{R}_{p, \mathbf{a}}$=$\mathbf{\Theta}_{3\times3}$ because by definition the interval between two consecutive B-spline knots has at most $q$+1 non-null coefficients, see \cite{Kress1998}. Using the stack vectors and matrices, the following linearized translational states transition equation is obtained in compact form
\begin{equation}
\mathbf{x_S}(r)=\mathbf{F}\mathbf{x}_r+\sum^{n_T}_{p=1}\left[\mathbf{G}_p(\mathbf{a_S})(\mathbf{u_S}_p+\Delta \mathbf{u}_{\mathbf{S}p})+\mathbf{H}_p(\mathbf{u_S}_p,\mathbf{a_S})\Delta\mathbf{a_S}\right], \label{state_propag_MPC}
\end{equation}
where the dependence with $r$ has been omitted at the right-hand side of Eq.\eqref{state_propag_MPC} for clarity. 

\subsection{Linearized planning problem}

The linearized planning problem seeks the same objectives as the NLP problem \eqref{opt_NLP} but now the terminal constraints are included in the cost function as in \cite{Gavilan2012} to encode a prescribed arrival time. This improves feasibility when considering disturbances, see \cite{Chen1998}, and can improve asymptotic stability properties without needing terminal constraints, see \cite{Limon2006}. At each MPC step, $r$, the linearized optimization problem to solve is 
\begin{equation}
\begin{aligned}
& \underset{\Delta \mathbf{u}_{\mathbf{S}p}, \Delta \mathbf{a_S}}{\text{minimize}}
& & \sum^{n_T}_{p=1}\sum^{r+N_p}_{i=r}\Delta u_{p,i} + \gamma_x \sum^{r}_{k=k_0} (\mathbf{x}_{N_p+k}-\mathbf{x}_f)^T\mathbf{Q}_x(\mathbf{x}_{N_p+k}-\mathbf{x}_f)\\ 
&&& + \gamma_v \sum^{r}_{k=k_0} \mathbf{x}^T_{N_p+k}\mathbf{Q}_v\mathbf{x}_{N_p+k}+\gamma_{\sigma}\sum^{r}_{k=k_0}(\pmb{\sigma}_{N_p+k}-\pmb{\sigma}_f)^T\mathbf{Id}(\pmb{\sigma}_{N_p+k}-\pmb{\sigma}_f)\\
&&&+\gamma_{\omega}T^2\sum^{r}_{k=k_0}\dot{\pmb{\sigma}}_{N_p+k}^T\mathbf{Id}\dot{\pmb{\sigma}}_{N_p+k}, \> \> \> \> k_0 = \text{max}(0,r-N_p),\\
& \text{subject to}
& & \mathbf{A}_{LS}\sum^{n_T}_{p=1}\mathbf{G}_p(r)(\mathbf{u_S}_p(r)+\Delta \mathbf{u}_{\mathbf{S}p}(r))+\mathbf{A}_{LS}\mathbf{H}(r)\Delta\mathbf{a_S}(r) \leq \\ 
&&&\mathbf{b}_{LS} - \mathbf{A}_{LS}\mathbf{F}(r)\mathbf{x}_r, \\
&&& \mathbf{0} \leq \mathbf{u_S}_p(r)+\Delta \mathbf{u}_{\mathbf{S}p}(r) \leq \mathbf{u}_{\mathbf{S}p,max}, \\
&&& \vert H_{i,rw}(t_{r+k,m},\mathbf{a_S}(r))+\Delta H_{i,rw}(t_{r+k,m},\mathbf{a_S}(r),\Delta\mathbf{a_S}(r))\vert \leq H_{i,max}, \\
&&& \vert \dot{H}_{i,rw}(t_{r+k,m},\mathbf{a_S}(r))+\Delta \dot{H}_{i,rw}(t_{r+k,m},\mathbf{a_S}(r), \Delta \mathbf{a_S}(r))\vert \leq \dot{H}_{i,max}, \\
&&& \Delta\pmb{\sigma}(t_r, \Delta\mathbf{a_S}(r)) = \hat{\pmb{\sigma}}_r-\pmb{\sigma}(t_r, \mathbf{a_S}(r)),\\
&&& \Delta\dot{\pmb{\sigma}}(t_r, \Delta \mathbf{a_S}(r)) = \dot{\hat{\pmb{\sigma}}}_r-\dot{\pmb{\sigma}}(t_r, \mathbf{a_S}(r)),\\
&&& \Delta\ddot{\pmb{\sigma}}(t_r, \Delta \mathbf{a_S}(r)) = \ddot{\hat{\pmb{\sigma}}}_r-\ddot{\pmb{\sigma}}(t_r, \mathbf{a_S}(r)),\\
&&& -\Delta \mathbf{u}_{\mathbf{S}p,max} \leq \Delta \mathbf{u}_{\mathbf{S}p} \leq \Delta \mathbf{u}_{\mathbf{S}p,max},\\
&&& -\Delta\mathbf{a}_{\mathbf{S},max} \leq \Delta \mathbf{a_S} \leq \Delta\mathbf{a}_{\mathbf{S},max}, 
\end{aligned}\label{opt_converted_lin}
\end{equation}
where the variables $\hat{\pmb{\sigma}}_r$, $\dot{\hat{\pmb{\sigma}}}_r$ and $\ddot{\hat{\pmb{\sigma}}}_r$ are the measured attitude and its derivatives at the end of each sampling interval. These values are necessary because the desired attitude path could suffer some deviations caused by reaction wheels saturation since the employed local reduction technique only guarantees time continuous constraint satisfaction at some discrete times. The increment on the reaction wheels kinetic momentum and its variation are
\begin{align}
\Delta H_{i,rw}(t_{r+k,m},\mathbf{a_S}(r),\Delta \mathbf{a_S}(r))=\sum^{r+n_c}_{j=r+1}\sum^{3}_{\alpha=1} \left. \frac{\partial H_{i,rw}}{\partial \sigma_\alpha}\frac{\partial \sigma_\alpha}{\partial a_{j,\alpha}} \right|_{\pmb{\sigma}_{r+k}, \mathbf{a}_j} \Delta a_{j,\alpha},\\
\Delta \dot{H}_{i,rw}(t_{r+k,m},\mathbf{a_S}(r),\Delta \mathbf{a_S}(r))=\sum^{r+n_c}_{j=r+1}\sum^{3}_{\alpha=1} \left. \frac{\partial \dot{H}_{i,rw}}{\partial \sigma_\alpha}\frac{\partial \sigma_\alpha}{\partial a_{j,\alpha}} \right|_{\pmb{\sigma}_{r+k}. \mathbf{a}_j} \Delta a_{j,\alpha},
\end{align}
The matrices associated with the terminal translational states costs terms are
\begin{equation}
\mathbf{Q}_{x} = 
\begin{bmatrix}
\mathbf{Id}_{3 \times 3} & \mathbf{\Theta}_{3 \times 3}\\
\mathbf{\Theta}_{3 \times 3} & \mathbf{\Theta}_{3 \times 3}\\
\end{bmatrix}, \> \> \> \> 
\mathbf{Q}_{v} = 
\begin{bmatrix}
\mathbf{\Theta}_{3 \times 3}  & \mathbf{\Theta}_{3 \times 3}\\
\mathbf{\Theta}_{3 \times 3} & \mathbf{Id}_{3 \times 3}\\
\end{bmatrix},
\end{equation}
and $\gamma_{x}$, $\gamma_{v}$, $\gamma_{\sigma}$ and $\gamma_{\omega}$ are positive scalars that weight the relative cost of each one of the terminal conditions with respect to fuel consumption. Since both the translational states propagation, see Eq.\eqref{state_propag_MPC}, and the flat output and its derivatives relation with the B-spline control points is linear, see Eq.\eqref{MRP_parameterised}, the proposed objective function is quadratic, hence the optimization problem \eqref{opt_converted_lin} is a QP problem.

\subsection{MPC scheme}

Summarizing the development of previous sections, the MPC scheme expressed as pseudocode is as follows
\begin{algorithm}[]
\Begin{Obtain a solution of the LP problem (\ref{opt_LP})\;
Transform the LP solution to NLP decision variables using Eq. (\ref{impulse_orientation})-(\ref{att_rot})\;
Obtain a solution of the NLP problem (\ref{opt_NLP}), $\mathbf{u_S}_p$ and $\mathbf{a_S}$\;
Apply $u_{p,0}$ and $\dot{\mathbf{H}}_{rw}(t)$ for $t$$\in$$[t_0, \> t_1)$\;
Initialize the current MPC step $r$=1\;
\While{$r \leq N_p$}{
Prescribe reference controls at $N_p+r$:\\
$u_{p,N_p+r}$=0,\\
$\pmb{\sigma}(t_{N_p+r},\mathbf{a_S}(r))$=$\pmb{\sigma}_f$,  $\dot{\pmb{\sigma}}(t_{N_p+r},\mathbf{a_S}(r))$=$\mathbf{0}$,  $\ddot{\pmb{\sigma}}(t_{N_p+r},\mathbf{a_S}(r))$=$\mathbf{0}$ $\longrightarrow$ $\mathbf{a_S}(r)$\;
Obtain the solution of the QP linearized rendezvous problem \ref{opt_converted_lin}, $\Delta \mathbf{u_S}_p(r)$ and $\Delta \mathbf{a_S}(r)$ \;
Update the decision variables $\mathbf{u_S}_p(r)$=$\mathbf{u_S}_p(r)$+$\Delta \mathbf{u_S}_p(r)$, $\mathbf{a_S}(r)$=$\mathbf{a_S}(r)$+$\Delta \mathbf{a_S}(r)$\;
Apply $\Delta V_{p,r}$ and $\dot{\mathbf{H}}_{rw}(t)$ for $t$$\in$$[t_r, \> t_{r+1})$\;
Update the current MPC step, $r$=$r$+1\;
}
}\caption{MPC scheme}
\end{algorithm}

The steps 2-4 are computed off-line while the vehicle is performing station-keeping around the departure point waiting the command to start the manoeuvre so no hard real-time requirements appear when computing this solution. However, the steps 8-14 within the while loop are performed on-line during the manoeuvre which require a fast computation. That is the main reason why a QP problem based on linearization around a previously computed solution has been developed instead of solving the NLP problem at each step. In \cite{Hartley2015} a field programmable gate arrays (FPGAs) implementation of a MPC scheme based on QP for elliptical orbits is shown to have the same computational performance as state of the art solvers.  

\section{Simulation results} \label{results}

Since the employed formulation does not make any assumptions on the chaser number of thrusters, two different scenarios will be considered. The first one will correspond to a heavy rendezvous satellite equipped with 10 thrusters while the other one corresponds to a low-power spacecraft equipped with 2 thrusters. The simulations of this sections have been obtained using MATLAB routines with \textit{Gurobi} optimization package, see \cite{Gurobi2014}, as LP and QP solver whereas the IPOPT optimization package, see \cite{IPOPT2016}, is used as NLP solver.

\subsection{Rendezvous model}

It is important to remark that although a linear model, see Eq.\eqref{xHCW}-\eqref{zHCW}, is used to compute the control sequence, the plant is considered to be dominated by the following non-linear relative motion dynamics, see \cite{Wie2008},
\begin{align}
\ddot{x}&=\ddot{\nu}z+2\dot{\nu}\dot{z}+\dot{\nu}^2x-\mu \frac{x}{[x^2+y^2+(r_t-z)^2]^{3/2}}, \label{xreal}\\
\ddot{y}&=-\mu \frac{y}{[x^2+y^2+(r_t-z)^2]^{3/2}}, \label{yreal}\\
\ddot{z}&=-\ddot{\nu}x-2\dot{\nu}\dot{x}+\dot{\nu}^2z-\mu \frac{z-r_t}{[x^2+y^2+(z-r_t)^2]^{3/2}}-\frac{\mu}{r_t^2}. \label{zreal}
\end{align}

\subsection{Disturbances model}

In a similar way as \cite{Gavilan2012} (note that there the disturbance is considered in an additive way), a disturbance on each of the thrusters performance is added to test the capabilities of the MPC scheme developed in Section \ref{MPC_scheme}. This disturbance is modelled in the chaser body axes as
\begin{equation}
\mathbf{u}^{B}_{p}(t_k)= \pmb{\Omega}(\pmb{\delta \theta}(t_k))\mathbf{w}_{p} u_p(t_k) (1+\delta u_p(t_k)), \> \> \> \> p=1\hdots n_T, \> \> \> \> k=1 \hdots N_p,
\end{equation}
where $u_p$ is the commanded output computed by the control laws, $\pmb{\delta \theta}\sim N_3(\text{E}[\pmb{\delta \theta}],$
$\pmb{\Sigma}_{\delta})$ is a vector of random small angles and $\delta u_p$$\sim N(\text{E}[\delta u_p], s[\delta u_p])$ is a random scalar. These disturbances model several physical aspects. First, the attitude control of the chaser will not be perfect, so one can expect some alignment errors, modelled by $\mathbf{\Omega}(\pmb{\delta \theta})$ in a simplified way. On the other hand, with $\delta u_p$ one can model thrust level disturbances.

\subsection{Simulation scenarios}

To test the capabilities of the proposed algorithm, two scenarios for different pursuer architectures are considered. For the first scenario, a heavy spacecraft equipped with 10 thrusters is considered while for the second scenario a lightweight satellite with a limited propulsion plant, with only 2 thrusters is simulated.

\subsubsection{Controller parameters}

Regarding controller parameters, for both cases, the B-splines order is chosen to be quintic which is equivalent to take $q$=5 in Eq.(\ref{MRP_parameterised}). The discrete grids sizes, to evaluate the time continuous constraints, are chosen as $n_L$=2 and $n_M$=12, while The objective function weights are taken as $\gamma_x$=10, $\gamma_v$=5, $\gamma_{\sigma}$=2 and $\gamma_{\omega}$=1. The LOS parameters for both cases are $c_y$=$c_z$=1/$\tan(\pi/4)$ and $y_0$=$z_0$=2.5 m. 

\subsubsection{Satellite with 10 thrusters}

In this scenario, a conventional cargo satellite with 10 thrusters has to rendezvous with a target flying in an eccentric low Earth orbit with $e$=0.1, $h_p$=600 km and $\nu(t_0)$=$\pi$/4. Table \ref{thrusters_layout_heavy} shows the characteristics of the considered propulsive layout.
\begin{table}[]  
\centering
\begin{tabular}{|c|c|c|c|c|c|}
\cline{1-6}
$p$ & $w_p$ & $u_{p,max}$ [m/s] & $p$ & $w_p$ & $u_{p,max}$ [m/s]\\ \hline
1 & [1, 0, 0]$^T$ & 1 & 6 & [0, 0, -1]$^T$ & 1\\ \hline
2 & [-1, 0, 0]$^T$ & 1 & 7 & [$\sqrt{2}$, $\sqrt{2}$, 0]$^T$/2 & 1\\ \hline
3 & [0, 1, 0]$^T$ & 1 & 8 & [$\sqrt{2}$, -$\sqrt{2}$, 0]$^T$/2 & 1\\ \hline
4 & [0, -1, 0]$^T$ & 1 & 9 & [-$\sqrt{2}$, $\sqrt{2}$, 0]$^T$/2 & 1\\ \hline
5 & [0, 0, 1]$^T$ & 1 & 10 & [-$\sqrt{2}$, -$\sqrt{2}$, 0]$^T$/2 & 1\\ \hline
\end{tabular}
\caption{Thrusters configuration for scenario 1}
\label{thrusters_layout_heavy}
\end{table}
On the other hand, the chaser inertia matrix is chosen to be similar to the russian Progress cargo spacecraft, see \cite{Fehse2003},
\begin{equation}
\mathbf{I}=
\begin{bmatrix}
31 & 0 & 0\\
0 & 31 & 0\\
0 & 0 & 5\\
\end{bmatrix}\cdot 10^3 \> \> \text{kg} \cdot \text{m}^2,
\end{equation}
whereas the bounds of the reaction wheels angular momentum and its variation are taken as $H_{i,max}$=500 N$\cdot$m$\cdot$s and $\dot{H}_{i,max}$=20 N$\cdot$m, respectively. At the beginning, the angular momentum of the system is considered to be null $\mathbf{H}_{tot}$=$\mathbf{0}$. The manoeuvre boundary conditions are given by Table \ref{boundary_conditions_heavy}. The considered disturbance parameters for this simulation are $\bar{\pmb{\delta}}$=0.0175, $\Sigma_{\delta,ij}$=0.0175$\delta_{ij}$, $\delta \bar{u}_p$=0.02 and $\sigma_{\delta u_p}$=0.05. 

\begin{table}[]  
\centering
\begin{tabular}{|c|c|c|c|}
\cline{1-4}
\multicolumn{4}{|c|}{Boundary conditions (intrinsic Euler angles sequence 3$\rightarrow$1$\rightarrow$3)} \\ \hline
$t_0$ & 0 s & $t_f$ & 900 s \\ \hline
$\mathbf{r}_0$ & [400, -250, -200]$^T$ m & $\mathbf{r}_f$ & [2, 0, 0]$^T$ m \\ \hline
$\mathbf{v}_0$ & [1, 1, -1]$^T$ m/s & $\mathbf{v}_f$ & [0, 0, 0]$^T$ m/s\\ \hline
$\pmb{\theta}_0$ & [0$^{\circ}$, 0$^{\circ}$, 0$^{\circ}$]$^T$ & $\pmb{\theta}_f$ & [90$^{\circ}$, 90$^{\circ}$, 90$^{\circ}$]$^T$\\ \hline
$\pmb{\omega}_{B/L,0}$ & [0$^{\circ}$, 0$^{\circ}$, 0$^{\circ}$]$^T$ s$^{-1}$ & $\pmb{\omega}_{B/L,f}$ & [0$^{\circ}$, 0$^{\circ}$, 0$^{\circ}$]$^T$ s$^{-1}$ \\ \hline
\end{tabular}
\caption{Scenario 1 boundary conditions}
\label{boundary_conditions_heavy}
\end{table}

\subsubsection{Satellite with 2 thrusters}  

In this scenario, a lightweight satellite with only 2 available thrusters has to rendezvous with a target flying in an eccentric low Earth orbit with $e$=0.5, $h_p$=400 $km$ and $\nu(t_0)$=$\pi$. The thrusters are mounted in an orthogonal configuration as shown by Table \ref{thrusters_layout_light}.

\begin{table}[]  
\centering
\begin{tabular}{|c|c|c|c|c|c|}
\cline{1-6}
$p$ & $w_p$ & $u_{p,max}$ [m/s] & $p$ & $w_p$ & $u_{p,max}$ [m/s]\\ \hline
1 & [0, 0, -1]$^T$ & 0.5 & 2 & [-1, 0, 0]$^T$ & 0.5\\ \hline
\end{tabular}
\caption{Thrusters configuration for scenario 2}
\label{thrusters_layout_light}
\end{table}

For this case, the chaser inertia matrix is chosen to be the one corresponding to the CNES small satellite MYRIADE, see \cite{Louembet2009}
\begin{equation}
\mathbf{I}=
\begin{bmatrix}
40 & -3 & -0.5\\
-3 & 28 & -1\\
-0.5 & -1 & 45\\
\end{bmatrix} \text{kg} \cdot \text{m}^2,
\end{equation}
whereas the bounds of the reaction wheels angular momentum and its variation are taken as $H_{i,max}$=1 N$\cdot$m$\cdot$s and $\dot{H}_{i,max}$=0.05 N$\cdot$m. At the beginning the angular momentum of the system is considered to be null, $\mathbf{H}_{tot}$=$\mathbf{0}$. In this case, the manoeuvre boundary conditions are shown in Table \ref{boundary_conditions_light} The considered disturbance parameters for this simulation are $\bar{\pmb{\delta}}$=$\mathbf{0}$, $\Sigma_{\delta,ij}$=0.0175$\delta_{ij}$, $\delta \bar{u}_p$=0 and $\sigma_{\delta u_p}$=0.01.

\begin{table}[]  
\centering
\begin{tabular}{|c|c|c|c|}
\cline{1-4}
\multicolumn{4}{|c|}{Boundary conditions (intrinsic Euler angles sequence 3$\rightarrow$1$\rightarrow$3)} \\ \hline
$t_0$ & 0 s & $t_f$ & 900 s \\ \hline
$\mathbf{r}_0$ & [350, 200, 200]$^T$ m & $\mathbf{r}_f$ & [2, 0, 0]$^T$ m \\ \hline
$\mathbf{v}_0$ & [1, 1, -1]$^T$ m/s & $\mathbf{v}_f$ & [0, 0, 0]$^T$ m/s\\ \hline
$\pmb{\theta}_0$ & [0$^{\circ}$ ,0$^{\circ}$, 0$^{\circ}$]$^T$ & $\pmb{\theta}_f$ & [90$^{\circ}$, 90$^{\circ}$, 90$^{\circ}$]$^T$ $^{\circ}$\\ \hline
$\pmb{\omega}_{B/L,0}$ & [0$^{\circ}$, 0$^{\circ}$, 0$^{\circ}$]$^T$ s$^{-1}$ & $\pmb{\omega}_{B/L,f}$ & [0$^{\circ}$, 0$^{\circ}$, 0$^{\circ}$]$^T$ s$^{-1}$ \\ \hline
\end{tabular}
\caption{Scenario 2 boundary conditions}
\label{boundary_conditions_light}
\end{table}

\subsection{Simulation results}

For each scenario, 100 realizations for the chosen disturbance parameters are simulated. Then, the obtained results are shown and discussed.

\subsubsection{Scenario with 10 thrusters}

\begin{figure}[] 
\begin{center}
\includegraphics[width=12cm,height=10cm,keepaspectratio]{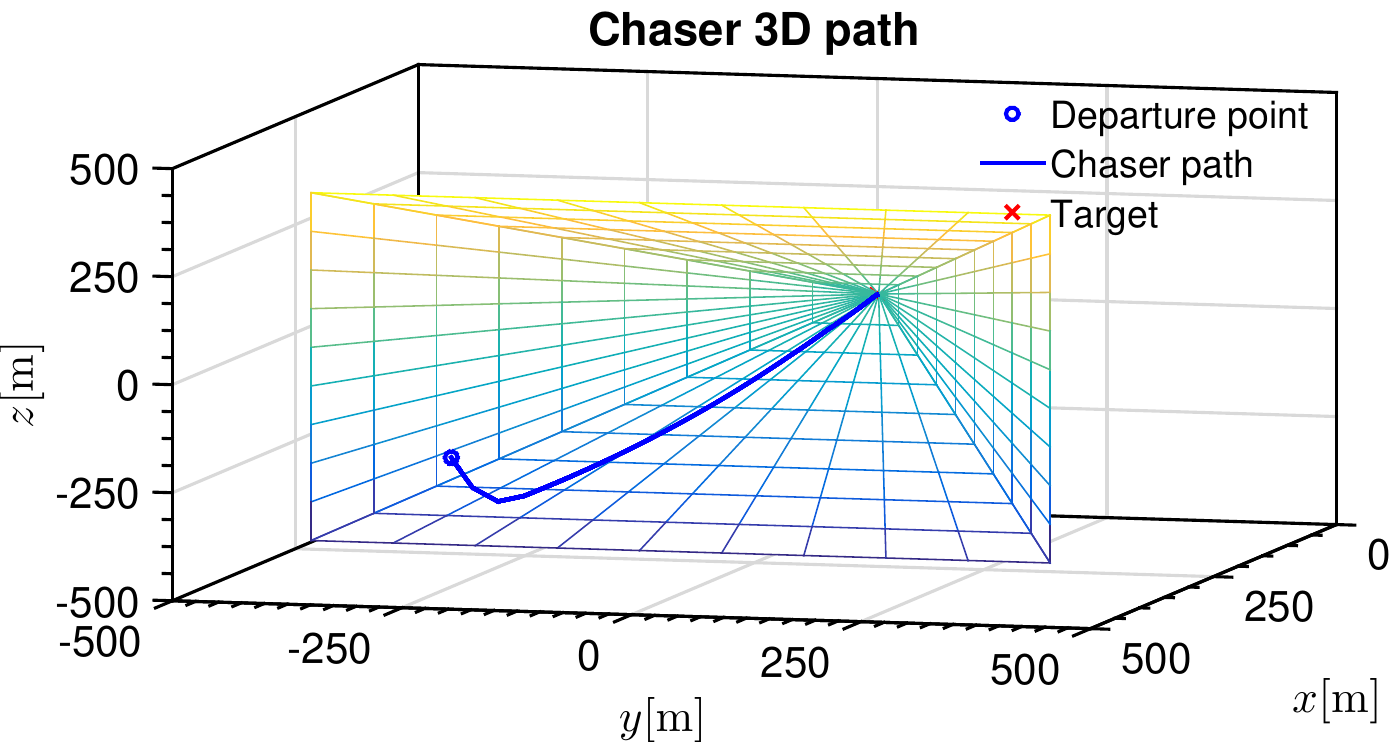}
\end{center}
\caption{Chaser 3D path of scenario 1 for the first random realization}
\label{3D_path_heavy}
\end{figure}

\begin{figure}[] 
\begin{center}
\includegraphics[width=12cm,height=10cm,keepaspectratio]{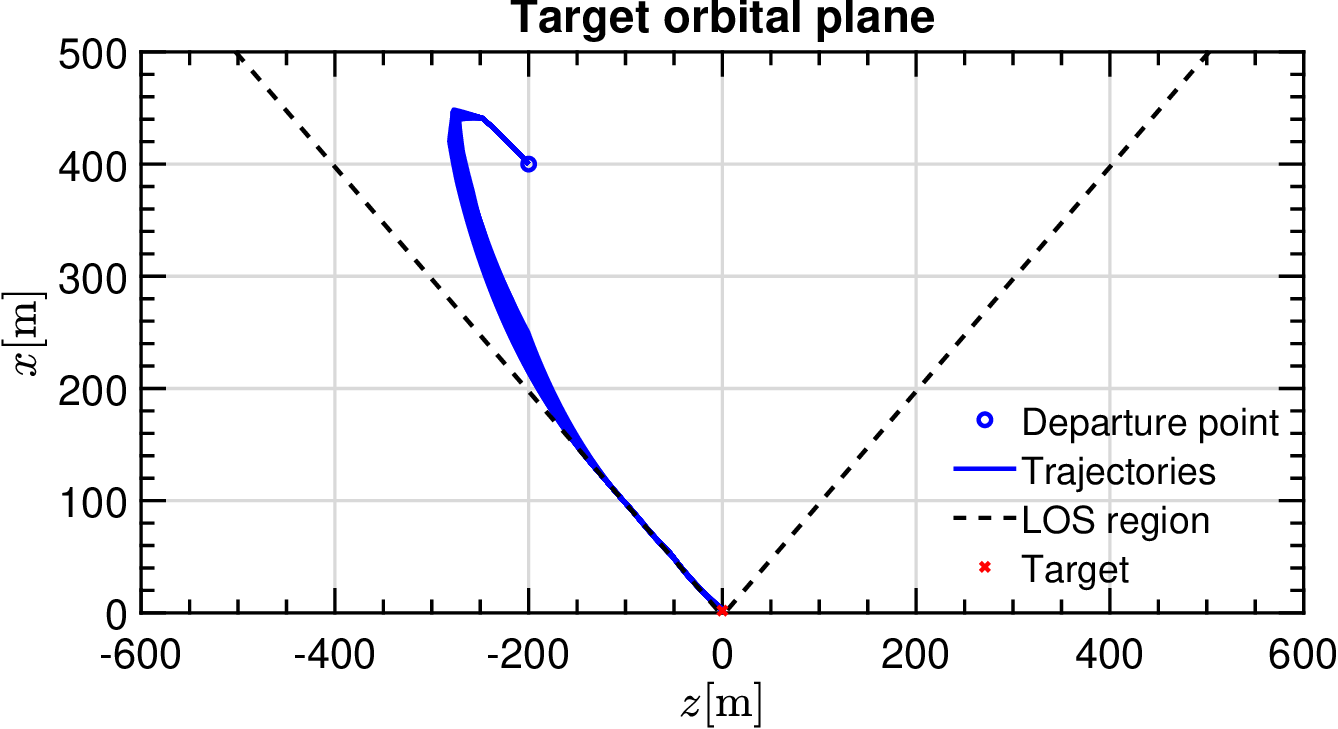}
\end{center}
\caption{Chaser trajectory on the target orbital plane of scenario 1 for all random realizations}
\label{XZ_plane_heavy}
\end{figure}

\begin{figure}[] 
\begin{center}
\includegraphics[width=12cm,height=10cm,keepaspectratio]{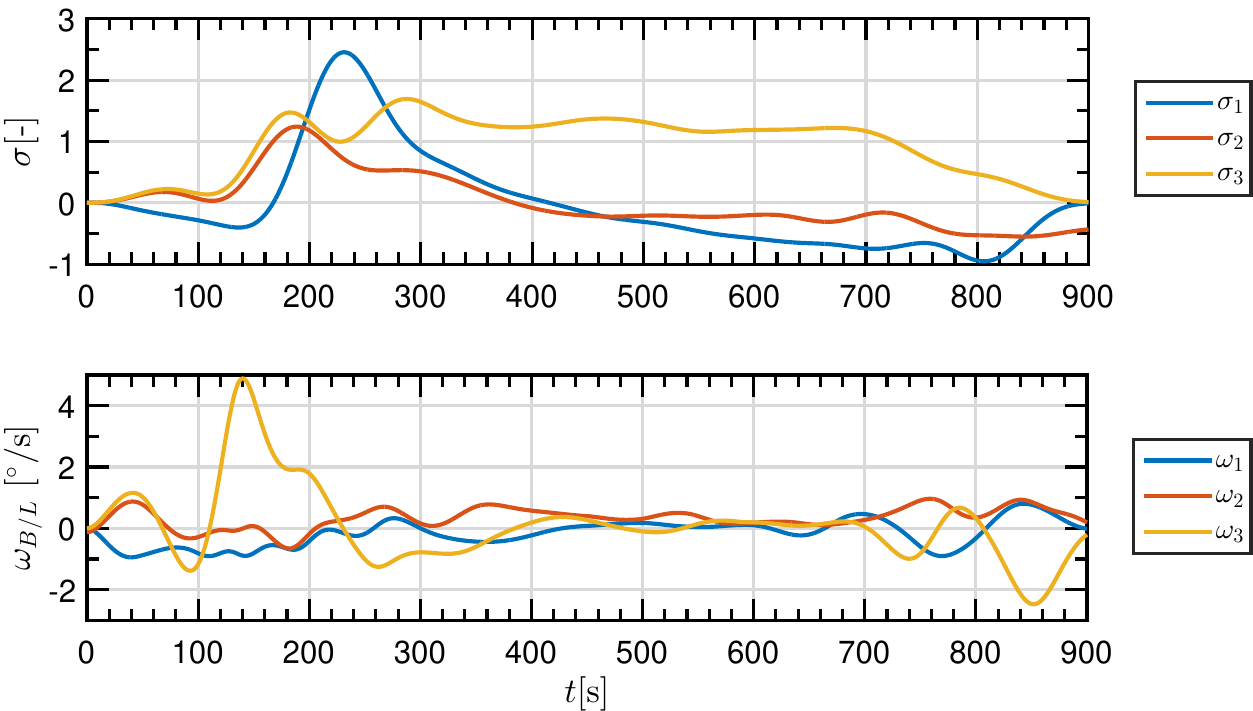}
\end{center}
\caption{Chaser attitude of scenario 1 for the first random realization}
\label{MRP_omega_evolution_heavy}
\end{figure}

\begin{figure}[] 
\begin{center}
\includegraphics[width=12cm,height=10cm,keepaspectratio]{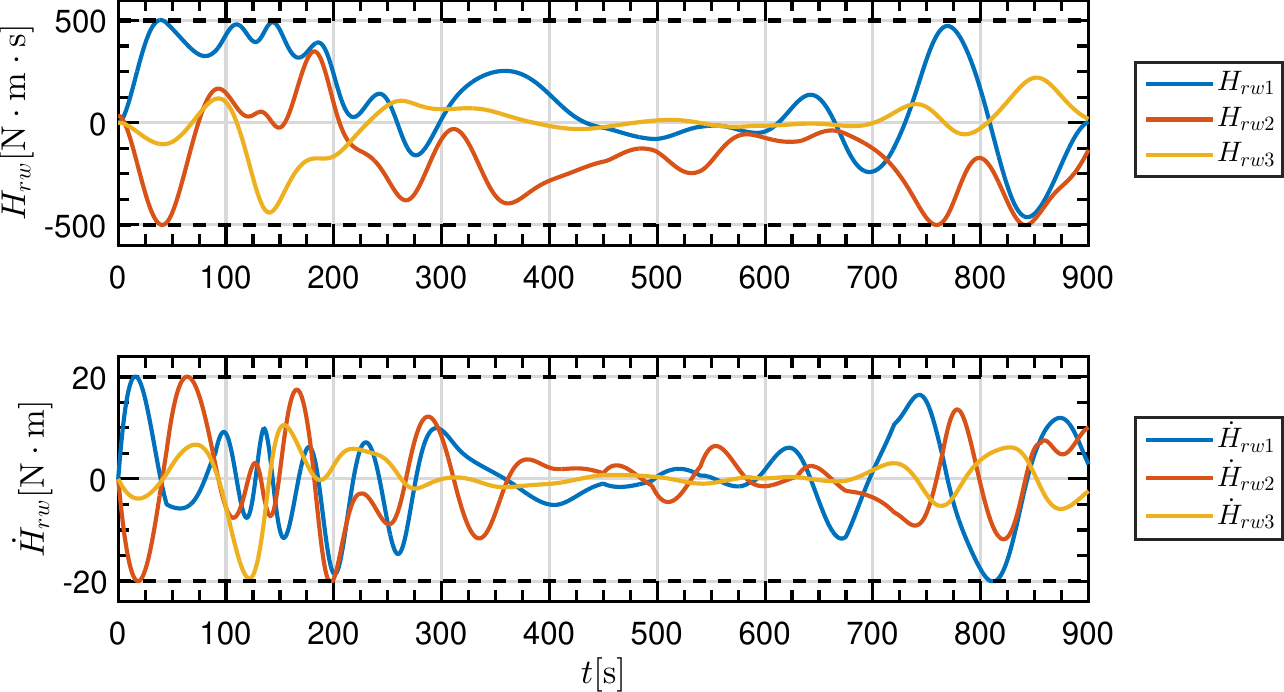}
\end{center}
\caption{Reaction wheels angular momentum and its variation of scenario 1 for the first random realization} 
\label{Reaction_wheels_heavy}
\end{figure}

First, analyse the scenario with 10 thrusters. For all the realizations, the linear QP program is feasible and the chaser reaches the proximity of the target without trespassing the LOS region, see Fig.\ref{XZ_plane_heavy}. A typical 3D path of a realisation is shown in Fig.\ref{3D_path_heavy} while the attitude profile is shown in Fig.\ref{MRP_omega_evolution_heavy}. For the shown realization the desired orientation is met at the end while the angular velocity is driven to a quasi-null value due to the considered uncertainties. More details on the terminal accuracy for this scenario are given in Table \ref{terminal_accuracy_heavy} where $\delta$ measures the mismatch between the obtained and the desired terminal value. Regarding the planned impulses, for the plotted realisation, see Fig.\ref{Impulses_light}, the thrusters \{1,2,6,9\} have relevant firings while thrusters \{3,4,5,7,8,10\} are not operated significantly along the manoeuvre. Regarding the cost, the NLP program reduces fuel consumption in a 21.054\% compared to the converted solution from the LP problem, see Table \ref{terminal_accuracy_heavy}. Finally, in Fig.\ref{Reaction_wheels_heavy} it is shown that the reaction wheels have saturations (both on angular velocity and acceleration) at the initial and final instants of the manoeuvre, but then desaturate immediately and keep providing torque.

\begin{figure}[] 
\begin{center}
\includegraphics[width=12cm,height=10cm]{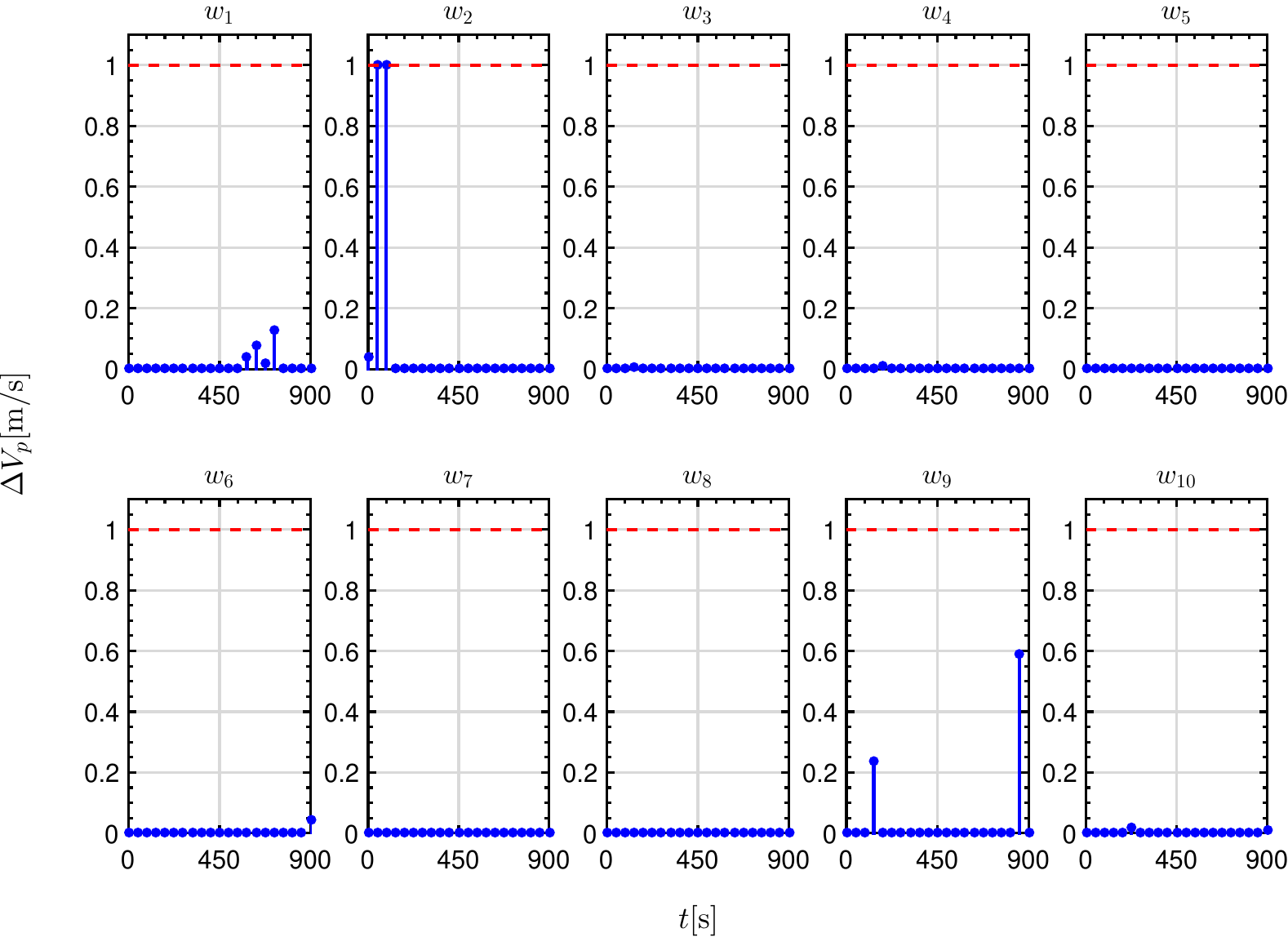}
\end{center}
\caption{Computed impulses of scenario 1 for the first random realization}
\label{Impulses_heavy}
\end{figure}

\begin{table}[]  
\centering
\begin{tabular}{|c|c|c|c|}
\cline{1-4}
\multicolumn{4}{|c|}{Results (intrinsic Euler angles sequence 3$\rightarrow$1$\rightarrow$3)} \\ \hline
\multicolumn{4}{|c|}{$J_{LP}$=4.08 m/s, $J_{NLP}$=3.15 m/s} \\ \hline 
\multicolumn{4}{|c|}{$\text{E} [J_{MPC}]$=3.24 m/s, $s [J_{MPC}]$=0.09 m/s} \\ \hline 
$\text{E}[\|\delta\mathbf{r}(t_f)\|_2]$ & 1.30 m & ${s}[\|\delta\mathbf{r}(t_f)\|_2]$ & 0.61 m \\ \hline $\text{E}[\|\delta\mathbf{v}(t_f)\|_2]$ & 2.80 cm/s & $s[\|\delta\mathbf{v}(t_f)\|_2]$ & 1.07 cm/s\\ \hline
$\text{E}$[$\pmb{\theta}(t_f)$] & [90.03$^{\circ}$, 93.97$^{\circ}$, 92.88$^{\circ}$] & 
$s$[$\pmb{\theta}(t_f)$] & [1.04$^{\circ}$, 1.45$^{\circ}$, 0.65$^{\circ}$]\\ \hline
$\text{E}[\|\pmb{\omega}(t_f)\|_2]$ & 0.31 $^{\circ}$/s & $s[\|\pmb{\omega}(t_f)\|_2]$ & 0.02 $^{\circ}$/s \\ \hline
\end{tabular}
\caption{Scenario 1 terminal results}
\label{terminal_accuracy_heavy}
\end{table}

\subsubsection{Scenario with 2 thrusters}

\begin{figure}[] 
\begin{center}
\includegraphics[width=12cm,height=10cm,keepaspectratio]{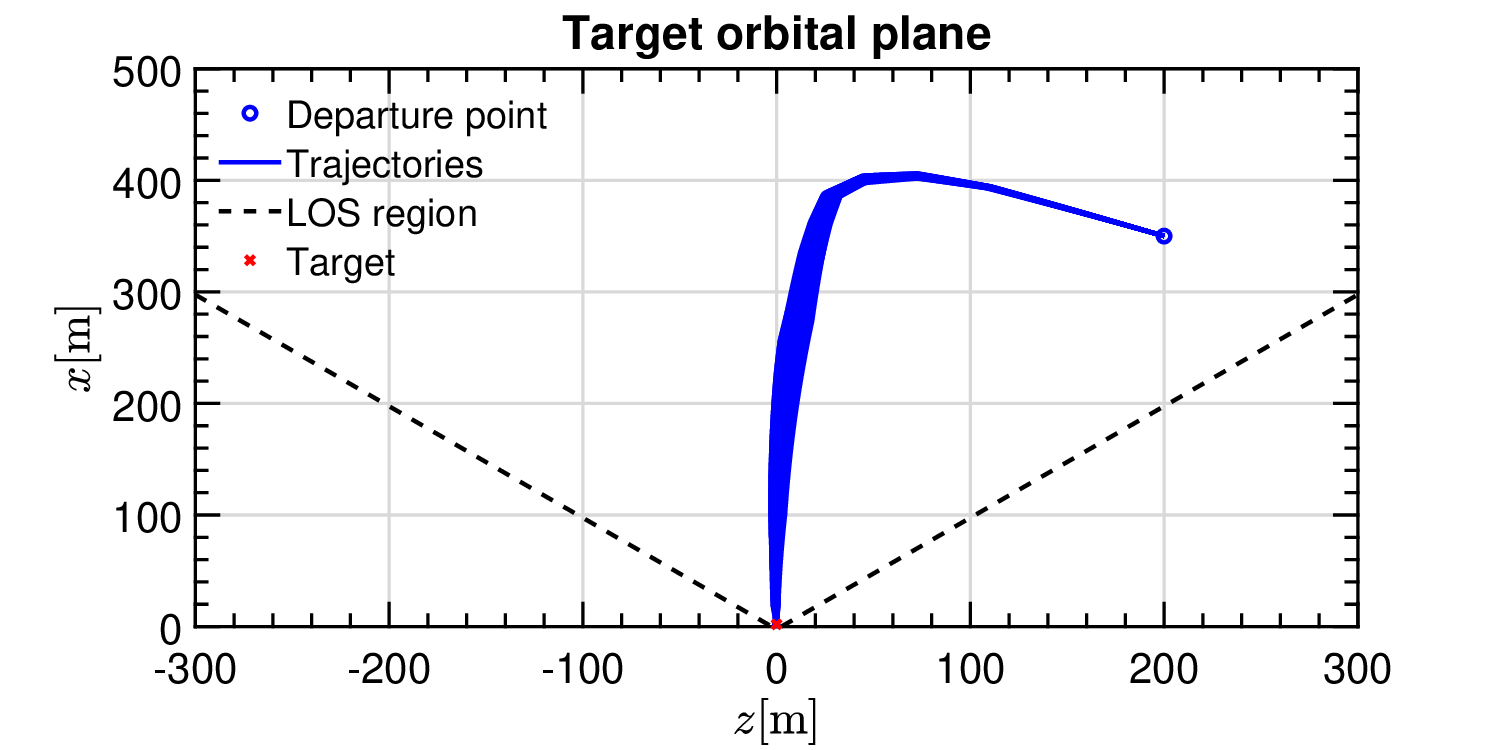}
\end{center}
\caption{Chaser trajectory on the target orbital plane of scenario 2 for all random realizations}
\label{XZ_plane_light}
\end{figure}

\begin{figure}[] 
\begin{center}
\includegraphics[width=12cm,height=10cm]{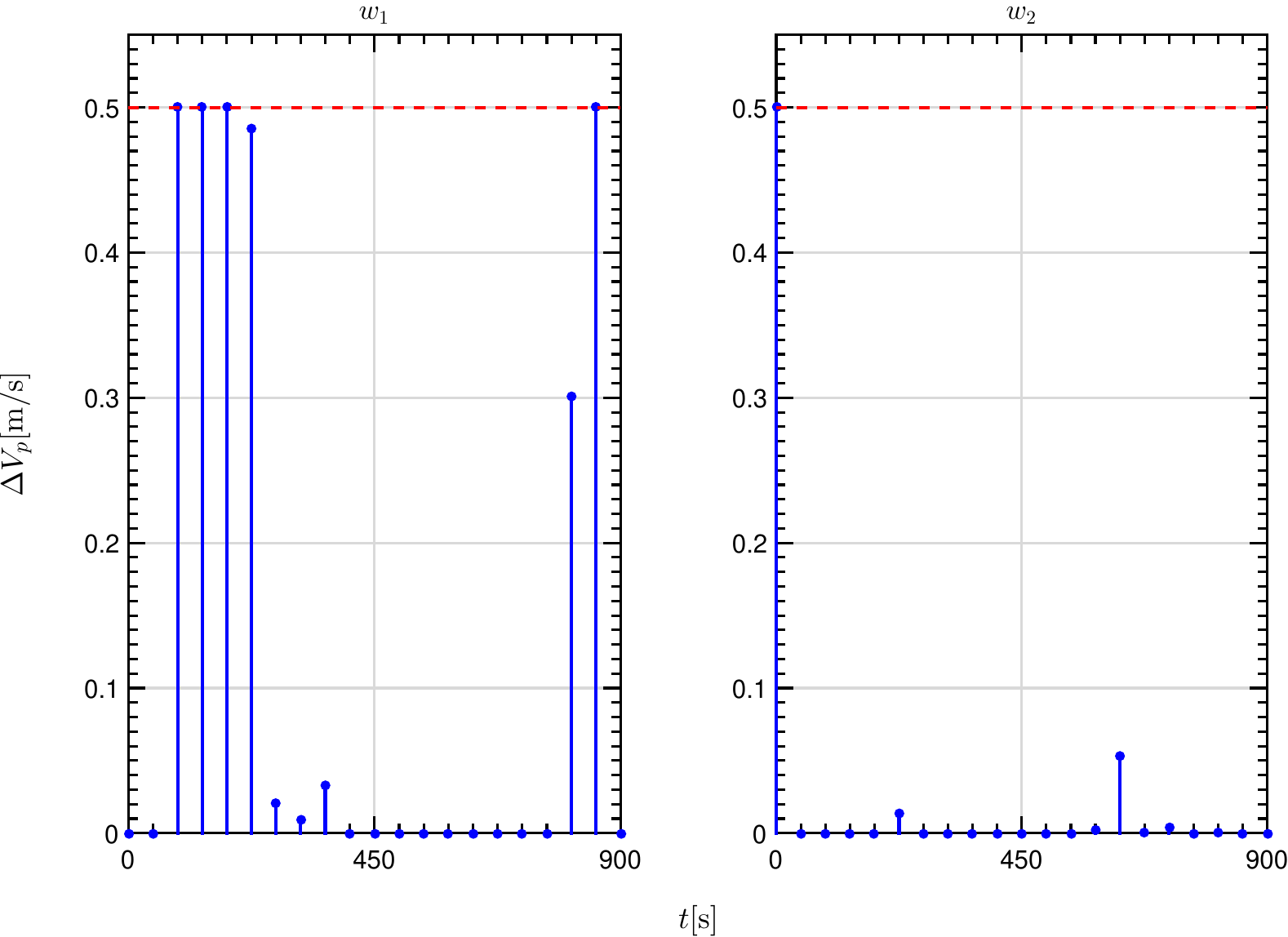}
\end{center}
\caption{Computed impulses of scenario 2 for the first random realization}
\label{Impulses_light}
\end{figure}

Analysing the second scenario with 2 thrusters, similar conclusions with the first scenario still holds, see Fig.\ref{XZ_plane_light}. Note that the spacecraft in this case is underactuated in translational control. Moreover, the desired final orientation is not favourable at all to brake the spacecraft since the thruster 1 nozzle will end pointing to the +$x$ axis and the thruster 2 nozzle to the -$z$ axis. The terminal accuracy is shown in Table \ref{terminal_accuracy_light} (it shows higher accuracy than the 10 thrusters scenario due to the lighter perturbations). For the plotted realization of Fig.\ref{Impulses_light}, it is shown that the final braking impulse has to be advanced one interval due to the non favourable last orientation. In this case, there is not an improvement in fuel consumption when compared to the obtained LP solution but the reaction wheels saturation peak has been lowered from 5.5608 N$\cdot$m$\cdot$s (LP solution converted to NLP solution) to 1 N$\cdot$m$\cdot$s, see Table \ref{terminal_accuracy_light}. 

\begin{figure}[] 
\begin{center}
\includegraphics[width=12cm,height=10cm,keepaspectratio]{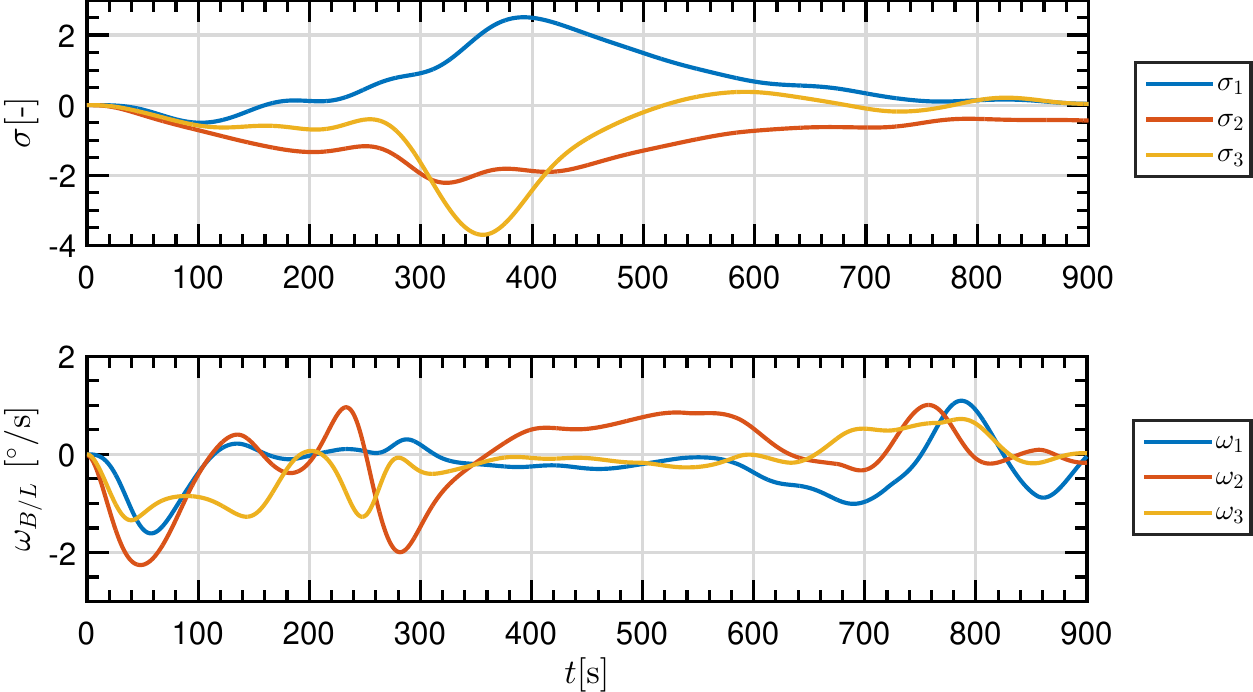}
\end{center}
\caption{Chaser attitude of scenario 2 for the first random realization}
\label{MRP_omega_evolution_light}
\end{figure}
\begin{figure}[] 
\begin{center}
\includegraphics[width=12cm,height=10cm,keepaspectratio]{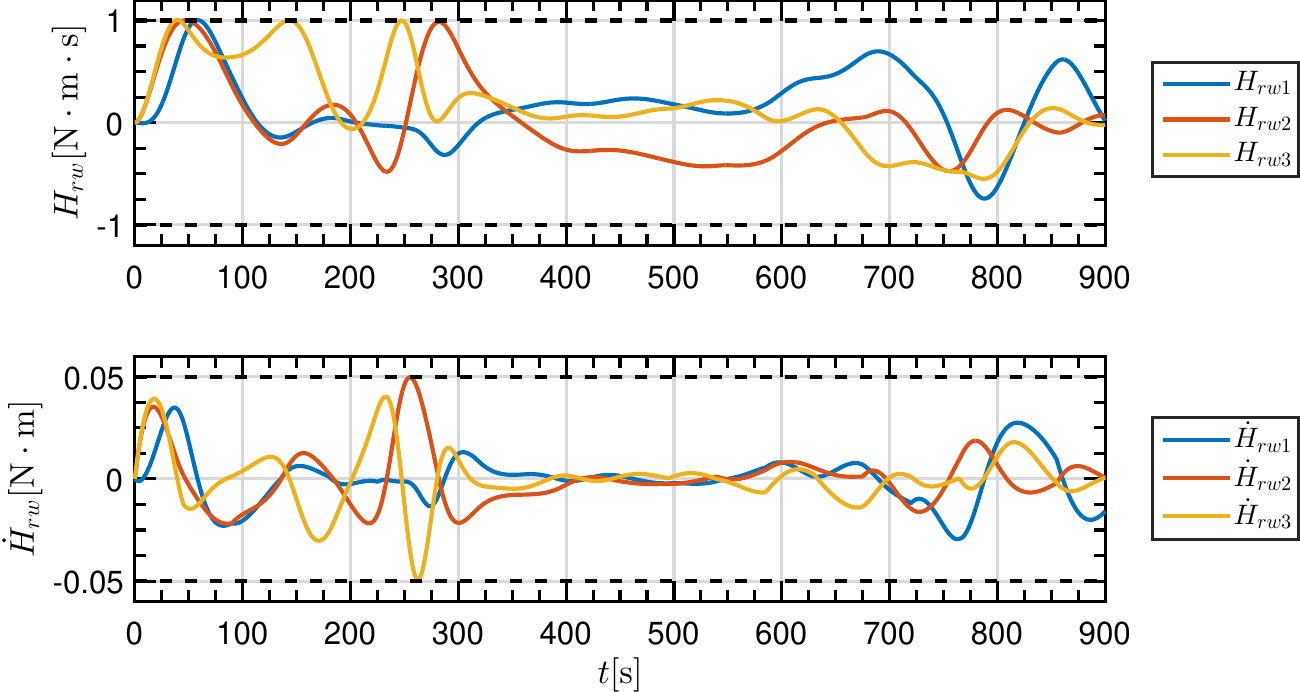}
\end{center}
\caption{Reaction wheels angular momentum and its variation of scenario 2 for the first random realization}
\label{Reaction_wheels_light}
\end{figure}

\begin{center}
\begin{table}[]  
	\centering
	\begin{tabular}{|c|c|c|c|}
		\cline{1-4}
		\multicolumn{4}{|c|}{Results (intrinsic Euler angles sequence 3$\rightarrow$1$\rightarrow$3)} \\ \hline
		\multicolumn{4}{|c|}{$J_{LP}$=3.49 m/s, $J_{NLP}$=3.40 m/s} \\ 
		\multicolumn{4}{|c|}{$\text{E} [J_{MPC}]$=3.43 m/s, $s [J_{MPC}]$=0.029 m/s} \\ \hline
	    $\text{E}[\|\delta \mathbf{r}(t_f)\|_2]$ & 0.82 m & ${s}[\|\delta \mathbf{r}(t_f)\|_2]$ & 0.36 m  \\ \hline
	    $\text{E}[\|\delta \mathbf{v}(t_f)\|_2]$ & 1.34 cm/s &  $s[\delta \mathbf{v}(t_f)]$ & 0.64 cm/s\\ \hline
		$\text{E}$[$\pmb{\theta}(t_f)$] & [97.39$^{\circ}$, 93.73$^{\circ}$, 89.10$^{\circ}$] & $s$[$\pmb{\theta}(t_f)$] & [4.57$^{\circ}$, 2.86$^{\circ}$, 2.51$^{\circ}$]\\ \hline
		$\text{E}[\|\pmb{\omega}(t_f)\|_2]$ & 0.31 $^{\circ}$/s & $s[\|\pmb{\omega}(t_f)\|_2]$ & 0.13 $^{\circ}$/s \\ \hline
	\end{tabular}
	\caption{Scenario 2 terminal results}
	\label{terminal_accuracy_light}
\end{table}
\end{center}
\section{Concluding remarks} \label{conclusions}

This paper has presented a predictive guidance and control algorithm for six-degrees of freedom spacecraft rendezvous based on the translational state transition matrix, the attitude flatness property, discretization and a MPC scheme based on linearization. One of the main contributions of the proposed algorithm is its ability to consider several chaser spacecraft configurations which not only reduces fuel consumption but also allows to consider propulsive and ACS constraints in an integrated framework.

The numerical experiments shown in Section \ref{results} have validated the method for two different spacecraft configurations. Additionally, the simulations have demonstrated convergence of the proposed MPC to the desired final state even in the presence of disturbances. However, a formal proof of stability has not been addressed and is left as future work.  

Possible future research lines include the following. First, it will be of great interest to consider on/off thrusters as it is done in \cite{Vazquez2017}. This will cause continuous coupling between translational and rotational motion since the vehicle will be spinning when the thrusters are fired. Second, to improve the robustness of the underactuated case, robust MPC techniques in the spirit of \cite{Gavilan2012} could be considered. Finally, another possible line is to consider more advanced techniques, that does not rely on discretization, to handle the time continuous constraints of the problem.

\section*{Acknowledgements}

The authors gratefully acknowledge Universidad de Sevilla for funding part of this work under its V-PPI US. Rafael Vazquez acknowledges financial support of the Spanish Ministerio de Econom\'ia y Competitividad under grant MTM2015-65608-P.

\bibliography{rendezvous_bib}

\end{document}